\documentclass[%
 aip,
 jmp,%
 amsmath,amssymb,
 reprint,%
]{revtex4-2}

\usepackage{graphicx}
\usepackage{dcolumn}
\usepackage{bm}
\usepackage{hyperref}
\usepackage{xcolor}
\usepackage{amsmath, amssymb}
\usepackage{tikz}
\usepackage{geometry}

\begin{document}

\preprint{AIP/123-QED}

\title{From Single-Particle Thermal Asymmetry to Collective Control: Janus Plasmonic Colloids for Photothermal Applications}

\author{Cédric Rousseau}
\affiliation{Micro- and Nanophotonic Materials Group, Research Institute for Materials Science and Engineering, University of Mons, Place du Parc 20, B-7000 Mons, Belgium}
\author{Javier González-Colsa}
\affiliation{Group of Optics, Department of Applied Physics, University of Cantabria, Avda. de los Castros s/n, 39005 Santander, Spain}
\author{Victor Joris}
\affiliation{Micro- and Nanophotonic Materials Group, Research Institute for Materials Science and Engineering, University of Mons, Place du Parc 20, B-7000 Mons, Belgium}
\author{Pablo Albella}
\affiliation{Group of Optics, Department of Applied Physics, University of Cantabria, Avda. de los Castros s/n, 39005 Santander, Spain}
\author{Bjorn Maes}
\affiliation{Micro- and Nanophotonic Materials Group, Research Institute for Materials Science and Engineering, University of Mons, Place du Parc 20, B-7000 Mons, Belgium}
\author{Gilles Rosolen}
\affiliation{Micro- and Nanophotonic Materials Group, Research Institute for Materials Science and Engineering, University of Mons, Place du Parc 20, B-7000 Mons, Belgium}

\date{\today}

\begin{abstract}
Janus nanoheaters provide a route to break the spatial symmetry of thermoplasmonic heating and to generate directionally biased thermal gradients at the nanoscale. However, in realistic photothermal scenarios, nanoheaters rarely act as isolated objects, but rather operate within colloidal assemblies where collective thermal and electromagnetic effects reshape the local temperature landscape. Here, we investigate how the intrinsic thermal asymmetry of a single Janus plasmonic nanoheater is transformed when moving from isolated particles to planar random arrays under continuous-wave illumination. Using coupled electromagnetic and steady-state heat-transfer simulations, we first quantify the finite spatial range over which the single-particle Janus signature remains detectable. We then show that neighboring nanoheaters introduce two competing effects: a nearly symmetric collective thermal background that screens and homogenizes the Janus factor, and short-range neighbor-induced asymmetries that partially compensate this screening when interparticle distances become comparable to the spatial extent of the single-particle thermal anisotropy. Finally, we derive a closed-form analytical model for the array-averaged Janus factor as a function of nanoparticle number and surface density. The model predicts the screening-dominated regime from single-particle quantities and identifies deviations associated with near-field neighbor-induced contributions at higher densities. These results demonstrate a central trade-off for biomedical photothermal applications in continuous-wave operation: collective heating is required to reach relevant temperature increments, but it progressively erases the directional Janus character by thermal homogenization. Our framework therefore provides a practical route to predict and rationally design the spatial temperature distribution in plasmonic assemblies of Janus particles.
\end{abstract}


\maketitle

\section{\label{intro}Introduction}

Over the past two decades, thermoplasmonics has evolved, fostering scientific advances with transformative impact in technology and biomedicine. Key contributions include solar energy harvesting, photocatalysis, diagnostics, and additive manufacturing~\cite{baffou_applications_2020}, to name a few. These advances are founded on the same concept, the Localized Surface Plasmon Resonance (LSPR)~\cite{maier_plasmonics_2007}. LSPRs are collective oscillations of the free-electron plasma confined within metallic nanoparticles (NPs) that can be resonantly excited by light. This resonance leads to intense and localized electromagnetic-field enhancement, accompanied by efficient light-to-heat conversion\cite{borah_coupled_2019}. Among the main advantages of LSPRs, the production of nanoscale temperature gradients stands out~\cite{baffou_thermoplasmonics_2017,govorov_generating_2007,khorashad_combining_2016}. It promotes a wide range of thermal phenomena, such as heat-driven chemical reactions~\cite{CHEHADI20199,XU2025100312}, nanoscale fluid motion~\cite{Roxworthy2014,Ciraulo2021}, and controlled thermal effects in biological tissues~\cite{Ruhoff2024}.

In this context, thermoplasmonic NPs have established themselves as powerful tools capable of delivering heat with unprecedented precision. However, in most cases, the structures responsible for generating these temperature gradients are homogeneous, which results in highly symmetric thermal fields. Moreover, since large-scale fabrication typically favors geometrically symmetric designs, the overall symmetry of the generated field is further enhanced~\cite{Baffou2010,Jauffred2019,GonzalezColsa2022Nanoheater}. Although such configurations are useful for many applications, this isotropy of the thermal field inherently limits the specificity of heat delivery, as the surrounding medium is heated more uniformly, reducing spatial selectivity.

Beyond this isotropy, recent studies have revealed that Janus NPs are a powerful strategy to break thermal symmetry at the nanoscale\cite{GonzalezColsa2022}. By integrating materials with contrasting thermal properties into a single architecture, these hybrid structures introduce multiple interfaces that break spatial symmetry and enable modulation of heat dissipation~\cite{OlartePlata2022,GonzalezColsa2023,Jiang2023}. These heterogeneous domains promote localized thermal gradients together with directional heat flow, which leads to spatially selective heating. This phenomenon, referred to as the thermal Janus effect, opens opportunities to improve heat delivery in photothermal therapies, potentially reducing overheating of healthy tissue and improving treatment efficacy in target cells.

While single structures can efficiently generate heat through LSPRs, the absolute amount of energy dissipated per particle is strongly limited by its nanoscale volume. As a consequence, the temperature increment provided by an isolated nanoheater is often insufficient to reach damage thresholds in photothermal therapy. This limitation is further exacerbated by constraints on the available optical power density. For instance, the typical maximum permissible exposure (MPE) for skin is about $0.33~\mathrm{W\,cm^{-2}}$ under continuous-wave illumination in the infrared~\cite{Li2018}. Under such limitations, exploiting thermal collective effects becomes crucial\cite{rousseau_concurrent_2024}.

Unfortunately, collective behavior not only amplifies the overall temperature rise compared to isolated structures, but also modifies its nature, making it considerably more complex. Among the most influential effects, electromagnetic interactions between neighboring structures, often through dipole-dipole coupling, can modify both the spectral position and magnitude of peak absorption. In highly aggregated colloids\cite{rousseau_aggregation_2026}, these interactions may hinder excitation of the dipolar response, promoting the emergence of higher-order modes or even broadband absorption~\cite{Abid2024,Wang2021}. Electromagnetic screening within densely packed colloids can also lead to inhomogeneous distributions of resistive losses, which translate into spatially non-uniform heat sources~\cite{Baffou2010}. Additionally, collective effects reduce thermal confinement. Since heat transfer is intrinsically passive and diffusive, overlapping temperature fields tend to smooth thermal gradients and reduce the localized response~\cite{Baffou2013}. These phenomena can therefore reduce both predictability and controllability of the thermal field.

In photothermal therapy, spatial selectivity largely relies on biofunctionalization strategies that promote NP accumulation and aggregation around cancerous cells while minimizing attachment to healthy tissue~\cite{Riaz2018}. This process naturally induces a three-dimensional distribution of NPs following the topology of the target cell membrane. In this context, Janus NPs offer an opportunity to partially recover thermal control and localization within colloidal assemblies. However, this opportunity comes with an inherent difficulty in continuous-wave operation. The collective heating needed to reach therapeutically relevant temperature increments also generates a smooth thermal background that can homogenize the temperature field and screen the directional Janus response. Understanding this balance between amplification of heating and loss of thermal directionality is therefore essential for assessing the practical relevance of Janus nanoheater assemblies in biomedical contexts.

Beyond photothermal therapy, directional heat release from Janus nanoheaters may also be useful in non-biomedical or \emph{ex vivo} applications, where higher irradiances can be applied. Examples include light-driven Janus micro/nanomotors based on self-thermophoretic propulsion~\cite{jiang_active_2010,liu_multiphysics_2023}. Recently, fuel-free and reversible self-thermophoretic propulsion was experimentally demonstrated for light-activated Au--SiO$_2$ Janus nanoparticles, with collective thermal effects also becoming relevant when several particles were simultaneously illuminated~\cite{truong_lightactivated_2026}. Related developments in plasmonic nanorobotics have shown that opto-thermophoretic forces can be exploited to capture, transport, reversibly assemble, and release bacteria, enabling light-driven nanorobots to operate as nanoscale robotic cleaners~\cite{qin_nanoscale_2026}. Other examples include photocatalytic Janus heterostructures~\cite{pradhan_janus_2009}, solar steam generation and desalination~\cite{wang_facile_2023}, and solar-assisted extraction from seawater~\cite{fu_janus_2023}. These examples highlight the broader interest of controlling anisotropic nanoscale heat release for propulsion, biological manipulation, and energy-related applications beyond hyperthermia.

Despite the growing interest in Janus nanoheaters, it remains unclear to what extent the thermal asymmetry of an isolated nanoheater survives when many particles collectively contribute to heat generation. Since practical photothermal scenarios inevitably involve clustered or aggregated NPs, understanding the competition between directional heating and collective thermal homogenization constitutes a key open question. To the best of our knowledge, a quantitative description of how the thermal Janus response evolves from isolated nanoheaters to random colloidal assemblies is still lacking.

In this work, we show that the thermal Janus response of an isolated nanoheater is progressively screened when moving from single-particle heating to random colloidal assemblies. Using coupled electromagnetic and steady-state heat-transfer simulations, we identify the finite spatial range over which the single-particle Janus signature remains detectable and show how this range controls the emergence of short-range neighbor-induced asymmetries. These local contributions compete with the nearly symmetric collective thermal background generated by the surrounding NPs, which drives the system toward thermal homogenization as the array size and density increase.

Building on this physical picture, we derive a closed-form analytical model for the array-averaged Janus factor as a function of NP number and surface density. The model captures the screening-dominated regime from single-particle properties and highlights the transition toward a denser regime where near-field neighbor-induced contributions become relevant. This framework clarifies why collective Janus heating remains possible over a finite range of interparticle distances, but also why direct biomedical implementation under continuous-wave illumination remains delicate when thermal homogenization dominates.

\section{\label{methods}Methods}

In this section, we present the numerical workflow employed to quantitatively assess the thermoplasmonic behavior of Janus nanoheater assemblies. We begin by introducing the thermal metric used throughout the manuscript, namely the Janus factor (Sec.~\ref{sec:janus_factor}). Subsequently, we describe the Janus monomer considered in this study and define the specific measurement points used to quantify the thermal asymmetry (Sec.~\ref{sec:monomer_description}). We then detail the procedure used to generate the planar nanoheater distributions (Sec.~\ref{sec:generation}) and summarize the COMSOL-based electromagnetic and thermoplasmonic simulations performed to compute the absorption cross sections and steady-state temperature fields, together with the optical and thermal properties employed in the simulations (Sec.~\ref{sec:comsol}). Finally, we introduce the analytical decomposition used to separate the intrinsic Janus response of a given nanoheater from the collective thermal background generated by the surrounding assembly (Sec.~\ref{sec:analytical_model_local}).

\subsection{Janus factor}
\label{sec:janus_factor}

To quantify the degree of thermal asymmetry generated by a system of NPs, we introduce a dimensionless parameter $F_J$, referred to as the thermal Janus factor. Let $T_1$ and $T_2$ be the temperatures in two opposite observation regions located on either side of the system, for instance front and rear regions, and let $T_{\mathrm{amb}}$ be the ambient temperature. The respective temperature increments are expressed as $\Delta T_1 = T_1 - T_{\mathrm{amb}}$ and $\Delta T_2 = T_2 - T_{\mathrm{amb}}$. Under this notation, the thermal Janus factor is defined as
\begin{equation}
\label{def_facteur_Janus}
F_J =
\frac{\Delta T_1 - \Delta T_2}
{\Delta T_1 + \Delta T_2}.
\end{equation}

From Eq.~\eqref{def_facteur_Janus}, it follows that $\lvert F_J\rvert \leq 1$, provided that $\Delta T_1$ and $\Delta T_2$ are positive. The sign of this factor indicates the directionality of the thermal response: positive values indicate that region 1 is hotter than region 2, whereas negative values correspond to the opposite situation. The magnitude of $F_J$ quantifies the quality of the thermal asymmetry, namely the temperature contrast between the two regions relative to the overall heating level. In this sense, $F_J$ does not measure the absolute temperature difference alone, but rather how efficiently the generated heat is distributed asymmetrically between both regions. Accordingly, $F_J=0$ corresponds to a thermally symmetric response, whereas $\lvert F_J\rvert \to 1$ corresponds to an increasingly ideal asymmetric response. As a result, adding a symmetric background heating contribution to both probe regions reduces $\lvert F_J\rvert$, even if the absolute temperature difference is unchanged.

Notably, this parameter is invariant to the incident irradiance, since a uniform change in excitation power rescales both $\Delta T_1$ and $\Delta T_2$ by the same factor. Similar formulations have been proposed in the literature to describe the concept of thermal inhomogeneity~\cite{Baffou2010}. Eq.~\eqref{def_facteur_Janus} is also related to ratio-based Janus metrics, such as $\zeta=\Delta T_1/\Delta T_2$~\cite{xie_regulating_2024}, through
\begin{equation}
F_J =
\frac{\zeta - 1}{\zeta + 1},
\qquad
\zeta =
\frac{1 + F_J}{1 - F_J}.
\end{equation}

The formulation proposed in Eq.~\eqref{def_facteur_Janus} is therefore more convenient than ratio-based metrics because it remains strictly bounded, whereas ratio-based metrics can become singular and less straightforward to interpret.

\subsection{Monomer Janus description}
\label{sec:monomer_description}

\begin{figure}[!ht]
 \centering
 \includegraphics[width=0.7\columnwidth]{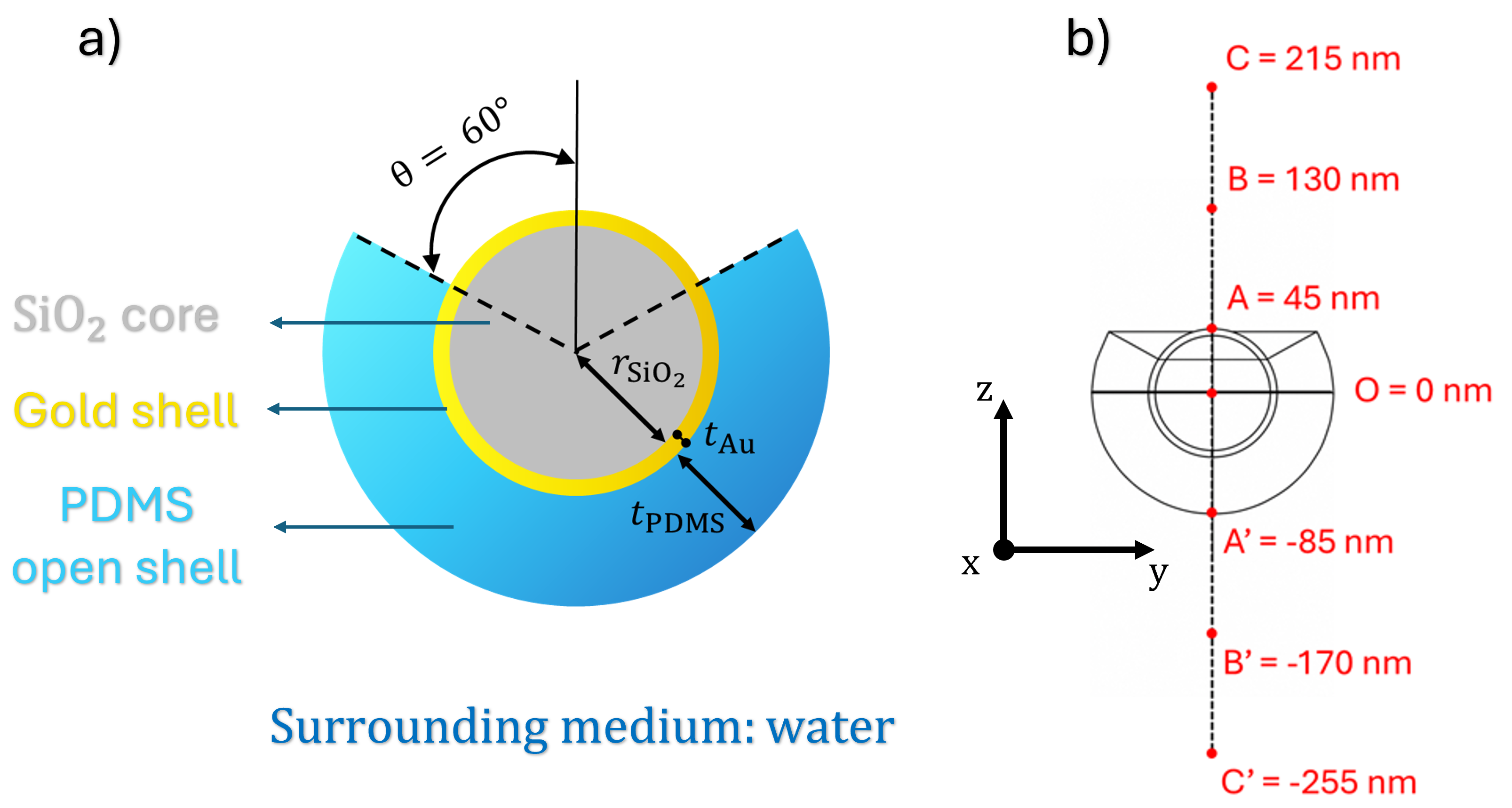}
 \caption{\label{monomer_schema}
 Schematics of the nanoheater geometry.
 (a) Cross-section of the Janus NP, consisting of a silica core of radius $r_{\mathrm{SiO_2}} = 40~\mathrm{nm}$ coated by a gold shell of thickness $t_{\mathrm{Au}} = 5~\mathrm{nm}$ and partially embedded in a PDMS domain of thickness $t_{\mathrm{PDMS}} = 40~\mathrm{nm}$. The opening angle $\theta$ defines the exposed cap, here $\theta=60^\circ$.
 (b) Definition of the temperature probe points, in red, placed along the NP symmetry axis (black dashed line) to evaluate the Janus factor from paired temperature readings.}
\end{figure}

The Janus monomer considered in this work, shown in cross-sectional view in Fig.~\ref{monomer_schema}(a), consists of a silica ($\mathrm{SiO_2}$) core of radius $r_{\mathrm{SiO_2}}=40~\mathrm{nm}$ surrounded by a continuous gold shell of thickness $t_{\mathrm{Au}}=5~\mathrm{nm}$. This metallic shell is further coated with an open polydimethylsiloxane (PDMS) overlayer of thickness $t_{\mathrm{PDMS}}=40~\mathrm{nm}$, characterized by an opening angle $\theta=60^\circ$. The NP is assumed to be immersed in water. This specific nanoheater configuration builds upon a recent study~\cite{gonzalezcolsa2026irisgated}, in which an optimal design space is identified for maximizing thermal directionality and focusing. The geometry selected for the present analysis falls within this optimized space.

A schematic representation of the probe points is shown in Fig.~\ref{monomer_schema}(b). The black dashed line denotes the radial symmetry axis of the NP, while the red dots indicate the probe points used to quantify the Janus effect. The pairs $(A, A')$, $(B, B')$, and $(C, C')$ are symmetrically positioned on opposite sides of the NP at different distances from the outer surface, namely 0, 85, and 170~nm, respectively. These sampling points allow for a spatially resolved assessment of how the thermal asymmetry decays into the surrounding medium. The open PDMS overlayer plays a central role in the thermal asymmetry because its low thermal conductivity locally hinders heat diffusion and redirects the heat flux toward the exposed metallic side of the particle.

\subsection{Generation of the different configurations}

\begin{figure}[!ht]
 \centering
 \includegraphics[width=1\columnwidth]{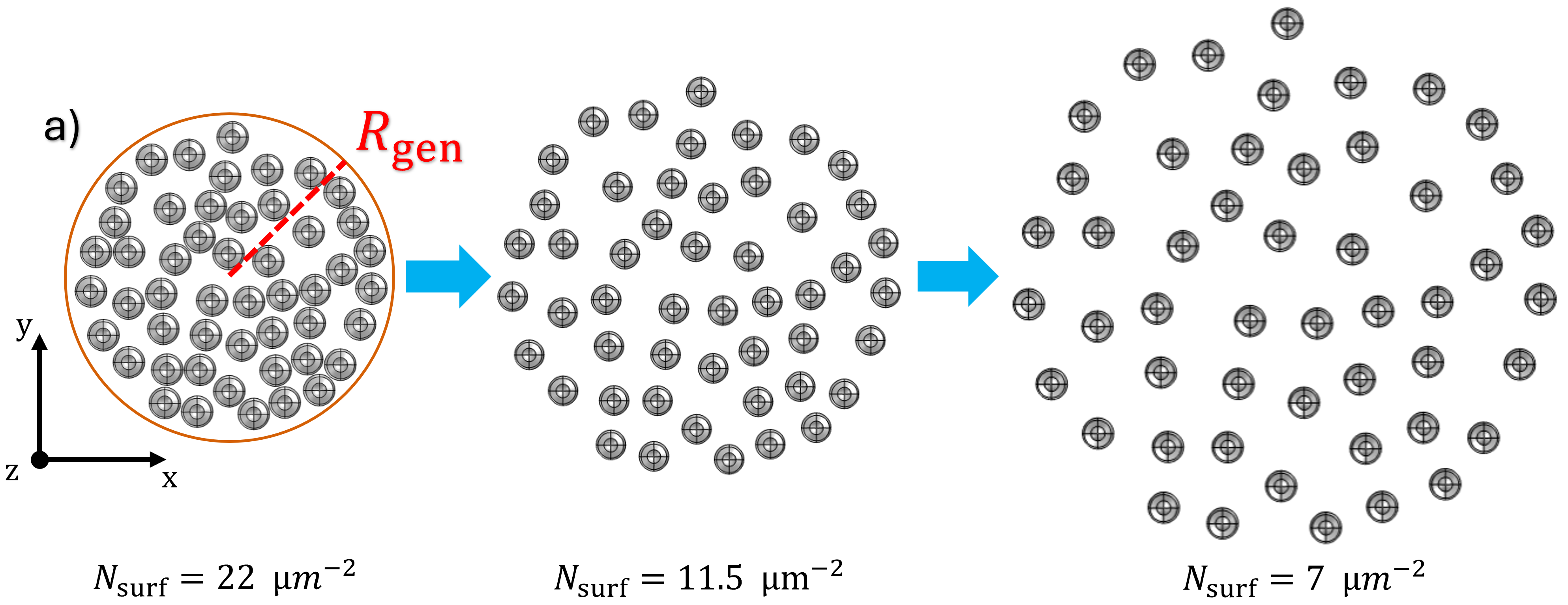}
 \caption{\label{full_geometry}
 Generation of planar NP arrays with controlled surface number density. Arrays with lower surface number densities, $N_{\mathrm{surf}}=11.5$ and $7~\mu\mathrm{m}^{-2}$, are obtained by radially stretching an initially dense configuration, $N_{\mathrm{surf}}=22~\mu\mathrm{m}^{-2}$. The stretching displaces each NP radially outward from the array center while preserving its azimuthal angle.}
\end{figure}

To assess the influence of collective effects, we generate planar NP arrays with controlled surface number densities $N_{\mathrm{surf}}$. To this end, a circular domain of radius $R_{\mathrm{gen}}$ containing NP assemblies with different $N_{\mathrm{surf}}$ was employed. We began with a dense planar configuration with $N_{\mathrm{surf}}=22~\mu\mathrm{m}^{-2}$, where $R_{\mathrm{gen}}$ is depicted with a dashed red line in Fig.~\ref{full_geometry}. This initial configuration was generated by randomly placing NPs under a non-overlap constraint that enforces a minimum edge-to-edge separation of 1~nm between neighboring NPs. In this initial state, all NPs share a common orientation, with their radial symmetry axis (black dashed line in Fig.~\ref{monomer_schema}(b)) perpendicular to the plane of the array. The side of the array where the NP openings are exposed is defined as the upper face of the array, while the opposite side is referred to as the lower face. The selected densities are intended to represent local nanoparticle-rich cell interfaces rather than uniform whole-tumor averages. Antibody-functionalized gold nanoshells have been shown to bind selectively to tumor-cell surfaces, while in vitro photothermal experiments have reported cellular loadings on the order of several thousand nanoshells per cancer cell, consistent with the order of magnitude of the local surface densities considered here~\cite{lowery2006immunonanoshells,stern2007efficacy}.
\label{sec:generation}
Lower concentrations were obtained by applying a uniform radial stretching to the initial configuration. From the center of the array, we attribute a radial distance $r$ and a polar angle $\phi_{\mathrm{plan}}$.  In these coordinates, each NP was displaced radially outward from the center of the array, increasing $r$, while preserving the azimuthal angle $\phi_{\mathrm{plan}}$. This procedure modifies the interparticle distance, and thus the concentration, without introducing additional structural disorder (see Fig.~\ref{full_geometry}). It therefore enables a direct comparison between assemblies of different densities derived from the same randomly generated initial distribution.

\subsection{COMSOL thermoplasmonic simulations}
\label{sec:comsol}

This study was performed using COMSOL Multiphysics v6.2, a finite element method solver that solves Maxwell's equations and the steady-state heat diffusion equation. The modeling follows a two-step workflow. First, the Electromagnetic Waves, Frequency Domain module was employed to solve Maxwell's equations and determine the electromagnetic response of the system under illumination. The excitation source was defined as a circularly polarized plane wave. In all simulations, the illumination direction was kept fixed and the incident wave impinged on the lower face of the array. This configuration represents a scenario in which a tumoral tissue is coated with NPs whose openings are oriented toward it. All simulations were performed at $\lambda=875~\mathrm{nm}$, which corresponds to the absorption resonance of the isolated Janus NP considered here\cite{gonzalezcolsa2026irisgated} and therefore provides a consistent reference wavelength for assessing both single-particle and collective photothermal responses.

The simulation domain was set as a spherical volume surrounded by perfectly matched layers (PMLs) to suppress spurious reflections and reduce artificial back-scattering. Gold was modeled as a lossy material using its complex dielectric function from Ref.~\onlinecite{johnson_optical_1972}. In contrast, $\mathrm{SiO_2}$, PDMS, and water were assumed to exhibit purely real permittivities, respectively $\varepsilon = 2.28$, $\varepsilon = 1.99$, and $\varepsilon = 1.77$, since their absorption is considered negligible over the considered wavelength range for the present geometry. The objective of this step is to calculate the spatial distribution of resistive losses, which serves as the volumetric heat source for the subsequent thermal simulation.

Importantly, electromagnetic interactions and optical shielding within the array lead to particle-dependent variations in the absorbed power. The absorbed power of the $i$th particle can be written as $P_{\mathrm{abs}}^{(i)}=I\sigma_{\mathrm{abs}}^{(i)}$, where $I$ is the incident irradiance and $\sigma_{\mathrm{abs}}^{(i)}$ is the absorption cross section of the considered NP within the array. As a result, the heat generation is not uniform across the assembly, but depends on the local electromagnetic environment of each particle.

In the second step, the Heat Transfer in Solids module was used to solve the steady-state heat diffusion equation,
\begin{equation}
\nabla \cdot \left(\kappa \nabla T \right) + Q_{\mathrm{abs}} = 0,
\end{equation}
where $\kappa$ is the thermal conductivity and $Q_{\mathrm{abs}}$ is the volumetric heat source obtained from the electromagnetic simulation. Heat transfer was assumed to be purely conductive, as the Rayleigh number in water remains sufficiently low to neglect convective effects under our conditions~\cite{baffou_thermoplasmonics_2017}. Radiative heat transfer was also considered negligible. The result is the temperature field throughout the domain, calculated from the resistive losses obtained in the first step.

A Dirichlet boundary condition was applied at the outer boundaries of the simulation region, fixing the temperature to $T_{\mathrm{amb}}=310~\mathrm{K}$. This fixed temperature defines the reference ambient temperature used to compute the temperature increments, $\Delta T = T - T_{\mathrm{amb}}$, in the Janus-factor analysis. Since such a condition may influence temperature gradients near the boundaries, infinite-element domains were employed to emulate an unbounded surrounding medium and ensure adequate separation between the array and the simulation boundaries. The thermal conductivities used in the simulations are $\kappa_{\mathrm{water}} = 0.6071~\mathrm{W\,m^{-1}\,K^{-1}}$, $\kappa_{\mathrm{PDMS}} = 0.15~\mathrm{W\,m^{-1}\,K^{-1}}$, $\kappa_{\mathrm{Au}} = 310~\mathrm{W\,m^{-1}\,K^{-1}}$, and $\kappa_{\mathrm{SiO_2}} = 1.3~\mathrm{W\,m^{-1}\,K^{-1}}$.

\subsection{Analytical model for the Janus factor of a NP within an assembly}
\label{sec:analytical_model_local}

To separate the intrinsic thermal asymmetry of a given Janus nanoheater from the collective thermal background generated by the surrounding assembly, we decompose the temperature rise at each probe point into two contributions\cite{baffou_thermoplasmonics_2017}. For a given probe pair of a specific NP in an array, namely $(A, A')$, $(B, B')$, or $(C, C')$, the total temperature rises measured in the array read
\begin{equation}
\Delta T_{1}^{\mathrm{tot}} =
\Delta T_{1}^{\mathrm{self}} + \Delta T_{1}^{\mathrm{col}},
\qquad
\Delta T_{2}^{\mathrm{tot}} =
\Delta T_{2}^{\mathrm{self}} + \Delta T_{2}^{\mathrm{col}},
\end{equation}
where $\Delta T_{1}^{\mathrm{self}}$ and $\Delta T_{2}^{\mathrm{self}}$ denote the temperature rises generated by the considered NP alone at the two probe points. The electromagnetic response is calculated for the complete array, while only the heat source associated with the considered NP is retained in the thermal simulation. Conversely, $\Delta T_{1}^{\mathrm{col}}$ and $\Delta T_{2}^{\mathrm{col}}$ denote the temperature rises generated at the same locations by all the other NPs, with the considered NP excluded from the thermal sources.

Using the Janus-factor definition introduced in Eq.~\ref{def_facteur_Janus}, the Janus response of a NP embedded in an assembly is therefore
\begin{equation}
\label{eq:FJ_array_local}
F_{J}^{\mathrm{array}} =
\frac{
(\Delta T_{1}^{\mathrm{self}} + \Delta T_{1}^{\mathrm{col}})
-
(\Delta T_{2}^{\mathrm{self}} + \Delta T_{2}^{\mathrm{col}})
}{
(\Delta T_{1}^{\mathrm{self}} + \Delta T_{1}^{\mathrm{col}})
+
(\Delta T_{2}^{\mathrm{self}} + \Delta T_{2}^{\mathrm{col}})
}.
\end{equation}

We then introduce three quantities: the isolated-particle Janus factor $F_{J0}$, the collective Janus factor $F_{J}^{\mathrm{col}}$, and the dimensionless ratio $\alpha$, defined as
\begin{equation}
\label{eq:FJ0_FJcoll_alpha}
F_{J0} =
\frac{\Delta T_{1}^{\mathrm{self}} - \Delta T_{2}^{\mathrm{self}}}
{\Delta T_{1}^{\mathrm{self}} + \Delta T_{2}^{\mathrm{self}}},
\qquad
F_{J}^{\mathrm{col}} =
\frac{\Delta T_{1}^{\mathrm{col}} - \Delta T_{2}^{\mathrm{col}}}
{\Delta T_{1}^{\mathrm{col}} + \Delta T_{2}^{\mathrm{col}}},
\qquad
\alpha =
\frac{\Delta T_{1}^{\mathrm{col}} + \Delta T_{2}^{\mathrm{col}}}
{\Delta T_{1}^{\mathrm{self}} + \Delta T_{2}^{\mathrm{self}}}.
\end{equation}

Using these definitions, Eq.~\ref{eq:FJ_array_local} can be rewritten as
\begin{equation}
\label{eq:FJ_array_decomp}
F_{J}^{\mathrm{array}} =
\frac{F_{J0}}{1+\alpha}
+
\frac{F_{J}^{\mathrm{col}}}{1 + 1/\alpha}.
\end{equation}

This expression provides a transparent physical decomposition of the measured Janus response. The first term corresponds to the intrinsic Janus factor of the isolated nanoheater, renormalized by the collective background heating. It is therefore associated with collective screening of the Janus response. The second term accounts for the asymmetry of the collective thermal field itself. In the nearest-neighbor configurations discussed below, this term can act as a positive near-field contribution when the neighboring particles are sufficiently close to create an asymmetric temperature rise at the two probe points.

The parameter $\alpha$ controls the balance between the self and collective contributions:when $0 < \alpha \ll 1$, the measured response is dominated by the isolated NP, whereas for $\alpha \gg 1$ the collective background progressively governs the Janus factor. It is worth noting that $\alpha$ plays a role analogous to the parameter introduced in Ref.~\onlinecite{Baffou2013} to distinguish between localized and collective heating regimes. In the following, \(F_J^{\mathrm{array}}\) will be used to denote the Janus factor whenever the considered NP is embedded in a system containing more than one NP.

In the low-density regime (from about $N_{\mathrm{surf}}=7~\mu\mathrm{m}^{-2}$ and below), the average interparticle separation is large. The collective thermal background generated by the surrounding NPs is therefore expected to vary only weakly between the two probe points of a given particle. The collective contribution can then be approximated as nearly symmetric,
$\Delta T_{1}^{\mathrm{col}} \simeq \Delta T_{2}^{\mathrm{col}} \equiv \Delta T_{\mathrm{col}}$,
and thus $F_{J}^{\mathrm{col}} \simeq 0$. Equation~\ref{eq:FJ_array_decomp} then reduces to
\begin{equation}
\label{eq:FJ_array_screening}
F_{J}^{\mathrm{array}} \simeq
\frac{F_{J0}}
{1 + 2\Delta T_{\mathrm{col}}/\left(\Delta T_{1}^{\mathrm{self}}+\Delta T_{2}^{\mathrm{self}}\right)}.
\end{equation}
This limiting form corresponds to a pure collective-screening regime, in which neighboring NPs primarily increase the overall heating level while reducing the thermal contrast of the considered particle.

\section{\label{results}Results}

This section presents the thermal response of Janus nanoheater assemblies. First, we analyze the temperature field generated by a single Janus NP (Sec.~\ref{sec:single_np}) and quantify the influence of nearest-neighbor interactions on the local Janus factor (Sec.~\ref{sec:nearest_neighbor}). Subsequently, we examine how these trends extend to disordered arrays of Janus NPs (Sec.~\ref{random_array}). Finally, we investigate the dependence of the array-averaged Janus factor on particle number and surface density, and compare the numerical results with the developed analytical model (Sec.~\ref{sec:array_averaged}).

\subsection{Thermoplasmonic response of a single Janus NP}
\label{sec:single_np}

\begin{figure}[!ht]
  \centering
  \includegraphics[width=0.9\columnwidth]{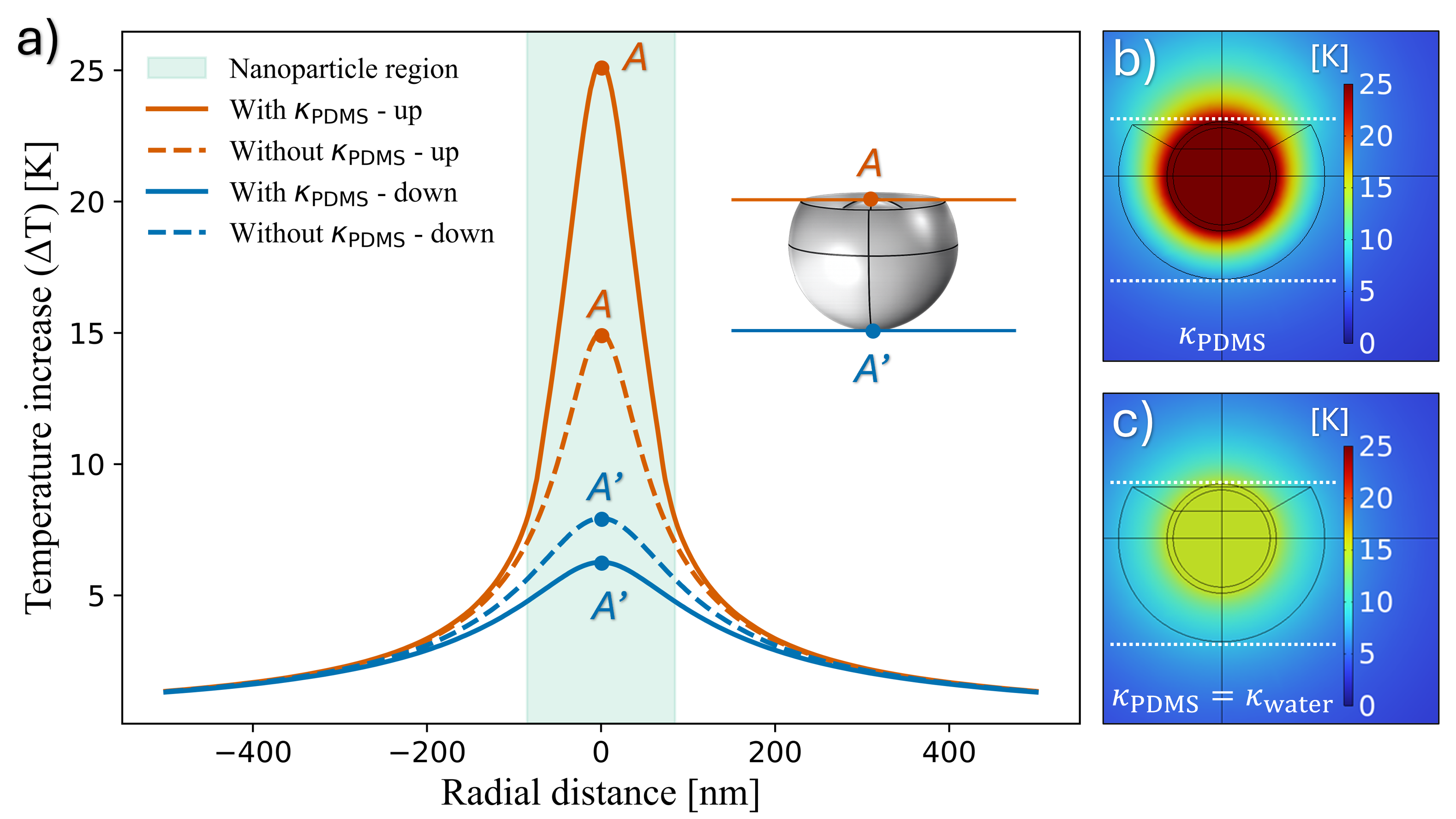}
  \caption{\label{Monomer_temp}
  Steady-state temperature increase $\Delta T$ generated by a single Janus NP under optical excitation. The irradiance is set to $1\times 10^{8}\ \mathrm{W,m^{-2}}$, and the calculated absorption cross section is $\sigma_{\mathrm{abs}} = 5.15\times 10^{-14}\ \mathrm{m^{2}}$.
  (a) Lateral profiles of $\Delta T$ extracted along the upper and lower probe lines, passing through points $A$ and $A'$, respectively. The orange and blue curves correspond to the upper and lower probe lines, respectively. Solid curves show the full case with PDMS thermal conductivity $\kappa_{\mathrm{PDMS}}$, whereas dashed curves show the reference case without PDMS thermal contrast, $\kappa_{\mathrm{PDMS}}=\kappa_{\mathrm{water}}$. The shaded area indicates the NP region along the scan.
  (b) Temperature map in the plane containing the two probe lines for the full case with PDMS thermal conductivity. White dashed lines mark the probe-line positions.
  (c) Temperature map in the same plane for the reference case without PDMS thermal contrast. White dashed lines mark the probe-line positions.}
\end{figure}

Fig.~\ref{Monomer_temp} illustrates the steady-state temperature increase generated by a single Janus NP. In Fig.~\ref{Monomer_temp}(a), the temperature profiles are extracted along two predefined horizontal probe lines located on the upper and lower sides of the particle, passing through the points $A$ and $A'$ (see Fig.~\ref{monomer_schema}(b)). Both lines are perpendicular to the particle symmetry axis (see inset in Fig.~\ref{Monomer_temp}(a)). These specific lines were selected to characterize the thermal environment at the exact locations where the measurement points of a neighboring Janus NP would be positioned when a second particle is introduced in the same plane and in close proximity to the first one. This provides a baseline for understanding how the local thermal field of a single NP influences the boundary conditions of adjacent structures in an assembly.

Two COMSOL simulations are compared using the same probe-line geometry. The first one corresponds to the full model, with the PDMS overlayer thermal conductivity set to $\kappa_{\mathrm{PDMS}}$ (solid lines in Fig.~\ref{Monomer_temp}). The second one suppresses the thermal-conductivity contrast of the PDMS overlayer by imposing $\kappa_{\mathrm{PDMS}}=\kappa_{\mathrm{water}}$ (dashed lines in Fig.~\ref{Monomer_temp}). In both cases, the optical properties of PDMS are kept unchanged so that the absorbed power remains the same. Thus, this comparison allows us to isolate the influence of the PDMS thermal conductivity on the steady-state thermal response of the NP.

The Janus nature of the isolated NP is first evidenced by comparing the full-model temperature profiles sampled along the upper and lower probe lines (solid orange and solid blue curves in Fig.~\ref{Monomer_temp}(a)). The temperature increase reaches $25.3$~K at point $A$, whereas it is only $6.3$~K at point $A'$. Using these values together with Eq.~\ref{def_facteur_Janus}, we obtain a Janus factor for an isolated monomer of $F_{J0}=0.60$. This thermal asymmetry is maximal in the immediate vicinity of the NP and progressively decays with radial distance. A non-negligible asymmetry persists up to approximately $300$~nm from the NP center, corresponding to about $215$~nm from its surface. For the specific nanoheater geometry considered here, this distance will be referred to as the characteristic thermal-Janus decay length. Beyond this distance, the upper and lower full-model profiles converge, and the Janus nanoheater effectively behaves as a point-like heat source.

The comparison between the full model (solid lines) and the reference case without PDMS thermal-conductivity contrast (dashed lines) clarifies the origin of this asymmetry. The presence of the PDMS thermal contrast induces a temperature decrease at point $A'$ and a corresponding increase at point $A$ (Fig.~\ref{Monomer_temp}(a)). This behavior is also visible from the comparison of the temperature maps in Figs.~\ref{Monomer_temp}(b) and~\ref{Monomer_temp}(c). It can be attributed to a redirection of the heat flux, as the low-conductivity PDMS acts as a thermal barrier that forces heat toward the exposed metallic surface, which was discussed in a previous work~\cite{gonzalezcolsa2026irisgated}. By evaluating the system without thermal contrast (dashed lines), namely $\Delta T=15.0$~K at point $A$ and $\Delta T=7.9$~K at point $A'$, we define a geometrical Janus factor for the isolated NP, $F^{\mathrm{geom}}_{J0}=0.31$. This baseline contribution arises solely from the spatial offset of the probe points relative to the NP center. Consequently, the overall Janus factor results from the combination of this pre-existing geometrical effect and the additional contribution provided by the thermal-conductivity contrast introduced by the PDMS overlayer.

As shown by the blue curves in Fig.~\ref{Monomer_temp}(a), the influence of the PDMS layer on the temperature rise taken along the lower probe line extends over the same distance range as the thermal asymmetry (the dashed and solid blue lines converge approximately 300~nm from the NP center). In contrast, the orange curves show that, along the upper probe line, the PDMS-induced effect remains confined to the immediate vicinity of the NP (the dashed and solid orange lines converge approximately 100~nm from the NP center), which is consistent with the larger PDMS coverage on the lower side of the nanoparticle. Importantly, for both probe lines relevant to our analysis, the influence of PDMS becomes negligible once the observation point exceeds the characteristic thermal-Janus decay length.

This loss of thermal asymmetry at long distances, and the resulting negligible impact of material thermal conductivity, has already been discussed in the literature~\cite{baffou_thermoplasmonics_2010,Baffou2010}. In our system, at distances exceeding the characteristic thermal-Janus decay length, the nanoheater operates in a regime where the temperature rise is governed by the total dissipated power rather than the internal distribution of heat sources. It therefore behaves as an effective isotropic point-like heat source, yielding a temperature rise that is essentially independent of both NP geometry and thermal conductivity.

\subsection{Nearest-neighbor effects on the Janus factor}
\label{sec:nearest_neighbor}

\begin{figure}[!ht]
  \centering
  \includegraphics[width=1\columnwidth]{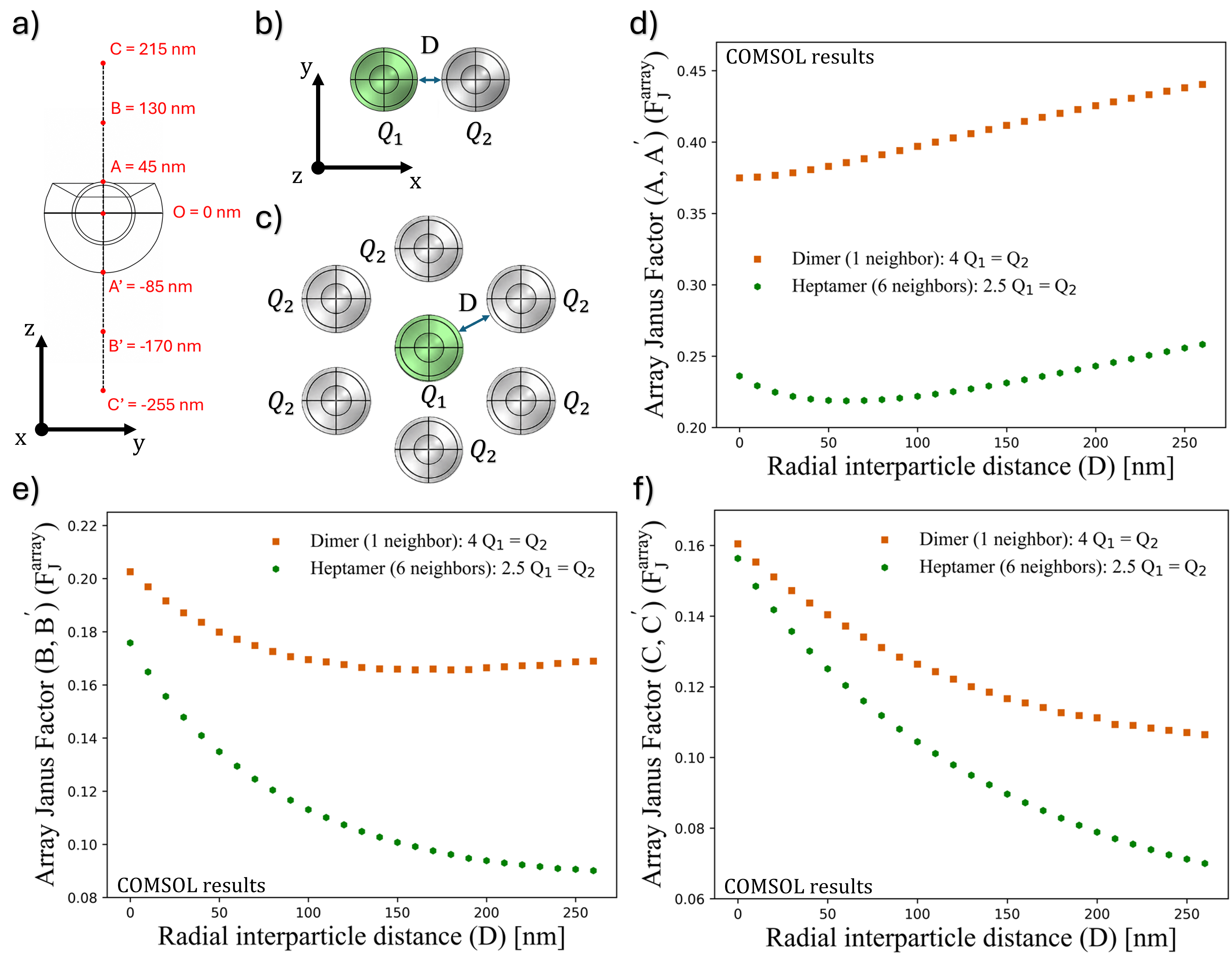}
  \caption{\label{Neighbour_contribution}
  Impact of neighboring NPs on the Janus factor as a function of the in-plane surface-to-surface separation D.
  (a) Definition of the probe pairs $(A, A')$, $(B, B')$, and $(C, C')$. (Same as Fig.~\ref{monomer_schema}(b))
  (b) One-neighbor configuration with heat rates $Q_1$ for the probed NP (green) and $Q_2$ for the neighboring NP (gray).
  (c) Six-neighbor configuration, where the probed NP is surrounded by six identical neighbors at the same separation D.
  (d) to (f) Janus factor calculated from (d) $(A, A')$, (e) $(B, B')$, and (f) $(C, C')$ for one neighbor with $Q_2=4Q_1$ (orange squares) and six neighbors with $Q_2=2.5Q_1$ (green dots).}
\end{figure}

To quantify how the local environment modifies the Janus response of a specific nanoheater, we examine two simplified planar configurations (Figs.~\ref{Neighbour_contribution}(b) and~\ref{Neighbour_contribution}(c)). The first is a dimer configuration (Fig.~\ref{Neighbour_contribution}(b)), where a single neighbor is positioned at an in-plane interparticle separation D from the probed NP. The second configuration is a planar heptamer (Fig.~\ref{Neighbour_contribution}(c)), where the probed NP is symmetrically surrounded by six identical neighbors at the same separation D. In both configurations, all NPs share the same orientation and D denotes the in-plane surface-to-surface separation between neighboring NPs. To explore various photothermal scenarios, we assign distinct volumetric heat generation rates, $Q_1$ for the central probed NP and $Q_2$ for the neighboring NPs. Specifically, for the dimer we use $Q_2 = 4\,Q_1$, for the heptamer we employ $Q_2 = 2.5\,Q_1$. These cases (where $Q_2 > Q_1$) are not intended to represent a unique experimental configuration, but rather to probe configurations in which neighboring particles act as stronger heat sources due to heterogeneous absorption within the array.

Figs.~\ref{Neighbour_contribution}(d) to~\ref{Neighbour_contribution}(f) show the Janus factor $F_J^{\mathrm{array}}$ of the green NP, as a function of the interparticle separation D, for the dimer and heptamer configurations (Figs.~\ref{Neighbour_contribution}(b) and (c)). This response is reported for the three probe pairs $(A, A')$, $(B, B')$, and $(C, C')$ (geometry repeated for convenience in Fig.~\ref{Neighbour_contribution}(a)), and for the particular heating ratios $Q_1/Q_2$. The graphs are calculated with COMSOL using both the full optical and thermal models.

The neighboring NPs modify the Janus response in a way that depends on both the observation distance (the particular point pairs) and the local heating environment (number of neighbors, separation D, heating ratio). In all investigated configurations, the Janus factor remains lower than that of the isolated monomer, \emph{i.e.}, $F_J^{\mathrm{array}}<F_{J0}=0.60$. Furthermore, the dependence on D changes markedly with the probe pair. Specifically, for the closest probe pair $(A, A')$, $F_J^{\mathrm{array}}$ generally increases with D (Fig.~\ref{Neighbour_contribution}(d)). For the intermediate pair $(B, B')$, the response becomes weakly non-monotonic for the dimer (orange points), and even decreasing for the heptamer (green points, Fig.~\ref{Neighbour_contribution}(e)). For the farthest pair $(C, C')$, $F_J^{\mathrm{array}}$ decreases for increasing D.

To understand these trends, we developed a planar point-source dimer model (described in detail in Sec.~S1 of the Supporting Information). In this model, each NP is reduced to a steady-state point heat source embedded in a homogeneous medium, and the Janus response is evaluated from the geometrical offset of the probe points with respect to the thermal sources. Therefore, the model captures mainly the geometrical contribution to the Janus response. It does not include the PDMS-mediated heat-flow redistribution (discussed for a single NP in Sec.~\ref{sec:single_np}). As a consequence, the Janus factors predicted by this point-source model remain lower than those obtained from the full COMSOL simulations. Nevertheless, the model provides a useful framework to interpret the competition between collective screening and neighbor-induced asymmetry, as discussed next.

\begin{figure}[!ht]
  \centering
  \includegraphics[width=0.85\columnwidth]{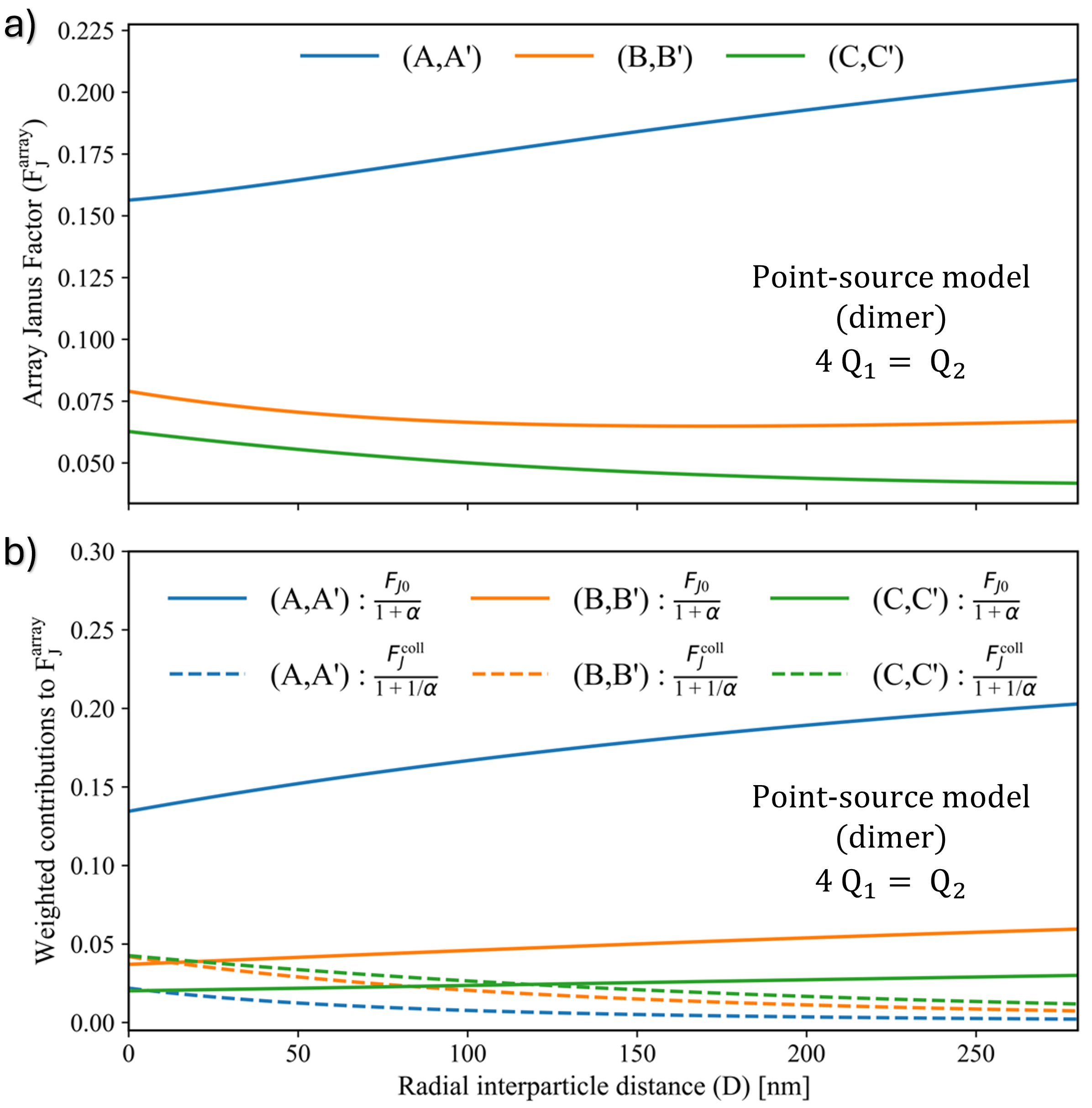}
  \caption{\label{fig:weighted_terms_main}
  Point-source model for the dimer case $Q_2=4Q_1$.
  (a) Janus factor $F_J^{\mathrm{array}}$ predicted by the point-source model as a function of the radial interparticle distance D (NP extension included in the model), for the three probe pairs $(A, A')$, $(B, B')$, and $(C, C')$.
  (b) Weighted contributions obtained by applying the decomposition of Eq.~\ref{eq:FJ_array_decomp} to the point-source model. Solid lines correspond to the screened self contribution, $F_{J0}/(1+\alpha)$, while dashed lines correspond to the neighbor-induced collective-asymmetry contribution, $F_J^{\mathrm{col}}/(1+1/\alpha)$. Curves with the same color correspond to the same probe pair.}
\end{figure}

Figure~\ref{fig:weighted_terms_main}(a) shows the Janus factor predicted by the point-source model for the dimer case $Q_2=4Q_1$. Despite its simplicity, the model is in good qualitative agreement with the COMSOL dimer calculations (orange squares in Figs.~\ref{Neighbour_contribution}(d) to (f)). Indeed, it reproduces the increase of $F_J^{\mathrm{array}}$ with D for the closest probe pair $(A, A')$ (Fig.~\ref{fig:weighted_terms_main}(a), blue curve), the weakly non-monotonic behavior for $(B, B')$ (Fig.~\ref{fig:weighted_terms_main}(a), orange curve), and the monotonic decrease for the farthest probe pair $(C, C')$ (Fig.~\ref{fig:weighted_terms_main}(a), green curve). This agreement indicates that the main dependence on interparticle separation is already captured by the geometrical imbalance between the probe points and the neighboring heat source.

With the point-source model, considering the superposition principle, we have full access to the temperature variations due to each source separately, and together. We can thus calculate all values of Eq.~\ref{eq:FJ0_FJcoll_alpha}, and determine the complete decomposition in Eq.~\ref{eq:FJ_array_decomp}, resulting in Fig.~\ref{fig:weighted_terms_main}(b). As mentioned, this decomposition separates the Janus factor into a (thermally) screened self contribution ($F_{J0}/(1+\alpha)$, solid lines in Fig.~\ref{fig:weighted_terms_main}(b)), and a collective-asymmetry contribution ($F_J^{\mathrm{col}}/(1+1/\alpha)$, dashed lines in Fig.~\ref{fig:weighted_terms_main}(b)).

We note that the collective Janus term ($F_J^{\mathrm{col}}/(1+1/\alpha)$, dashed lines in Fig.~\ref{fig:weighted_terms_main}(b)) is a near-field contribution. It becomes significant only when a neighboring NP is sufficiently close to the probed NP (small interparticle distance D) to induce an asymmetric temperature rise at the two measurement points. For the closest probe pair $(A, A')$, the screened self contribution remains dominant over the whole D distance range (blue solid line), whereas the near-field contribution remains comparatively small (blue dashed line). At further probe pairs, for $(B, B')$ and $(C, C')$, the asymmetry of the probed NP weakens (lower values for orange and green lines), and the relative importance of the near-field contribution increases (similar values for dashed and solid orange and green lines). This redistribution of weight between collective screening and near-field neighbor-induced asymmetry explains the different curve shapes observed in Figs.~\ref{Neighbour_contribution}(d) to~\ref{Neighbour_contribution}(f): The closest pair $(A, A')$ has an upward trend as a function of D (due to the dominating self-contribution, solid lines in Fig.~\ref{fig:weighted_terms_main}(b)), the farthest pair $(C, C')$ has a downward trend (due to the dominating collective-contribution, dashed lines in Fig.~\ref{fig:weighted_terms_main}(b)).
 
This interpretation is consistent with the single-particle analysis of Sec.~\ref{sec:single_np}, which showed that the asymmetric thermal signature of an isolated Janus NP extends only over the characteristic thermal-Janus decay length, approximately 215~nm from its surface for the geometry considered here. When the interparticle separation $D$ becomes comparable to this length, neighboring NPs can influence the thermal asymmetry of the probed NP and induce an additional asymmetric temperature rise at the measurement points. This produces a positive correction to the array Janus factor, which partially compensates the collective-screening effect. Conversely, when the interparticle separation $D$ becomes sufficiently large, this near-field contribution vanishes and the response progressively approaches a purely thermal screening-dominated regime.

In summary, these results demonstrate that the Janus factor of a non-isolated NP is not an intrinsic property of the NP alone. Rather, it is an extrinsic metric that depends on both the observation distance, through the choice of probe points, and the thermal background provided by neighboring particles. Unlike $F_{J0}$, which is fixed by the isolated NP geometry and material composition, $F_J^{\mathrm{array}}$ reflects the combined effect of the symmetric collective background, which screens the intrinsic Janus response, and short-range neighbor-induced thermal asymmetries, which become relevant when the interparticle spacing lies within the spatial range of the single-particle thermal gradient.

\subsection{Janus factor in randomly distributed arrays}
\label{random_array}

While the previous analysis focused on configurations with a well-defined local environment, more realistic scenarios typically involve a large number of NPs distributed without long-range order. In such disordered assemblies, each Janus NP is embedded in a heterogeneous thermal background arising from the collective heating of its neighbors, and the notion of a unique, intrinsic Janus response becomes insufficient. To address this more realistic scenario, we now investigate the behavior of the Janus factor in randomly distributed arrays containing multiple nanoheaters. This approach allows us to assess how positional disorder, surface number density, and probe distance jointly shape the local and statistical properties of the Janus factor.

\begin{figure}[!p]
  \centering
  \includegraphics[width=1\columnwidth]{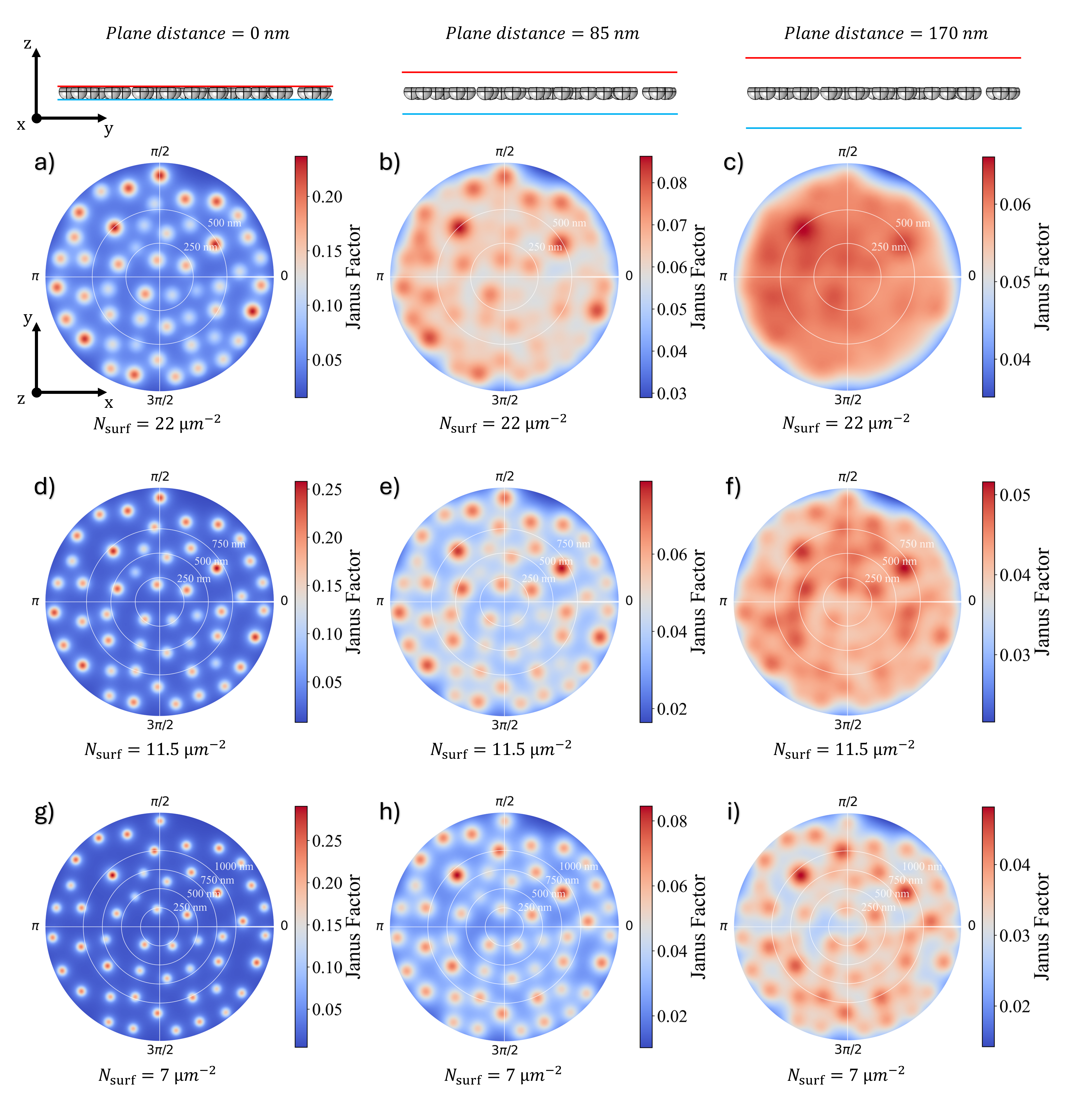}
  \caption{\label{Janus_map_planar}
  Spatial distribution of the local Janus factor (obtained with COMSOL) on a planar array for different surface number densities ($N_{\mathrm{surf}}$) and probe distances. Top row: side-view schematics indicating the evaluation planes located at probe distances of 0, 85, and 170~nm from the NP plane; red and blue lines denote the two probe planes used to compute the Janus factor. Panels (a) to (c), (d) to (f), and (g) to (i) show polar maps over a circular domain for $N_{\mathrm{surf}}=22$, 11.5, and $7~\mu\mathrm{m}^{-2}$, respectively. From left to right, columns correspond to probe distances of 0, 85, and 170~nm. The angular coordinate is the polar angle $\phi$, and the radial coordinate is the in-plane distance from the disk center, indicated by white circles. Color bars indicate the local Janus factor magnitude.}
\end{figure}

Figure~\ref{Janus_map_planar} shows spatial maps of the Janus factor, obtained with COMSOL, for planar, randomly distributed arrays of 50 Janus NPs (see Fig.~\ref{full_geometry}) at three probe distances, 0, 85, and 170~nm, and for three surface number densities, $N_{\mathrm{surf}}=22$, 11.5, and $7~\mu\mathrm{m}^{-2}$. The corresponding density-based horizontal particle-to-particle surface distance is estimated as $D_{\rho}\simeq N_{\mathrm{surf}}^{-1/2}-2R_{\mathrm{out}}$, where $R_{\mathrm{out}}\simeq 85~\mathrm{nm}$ is the effective outer radius of the Janus NP including the PDMS overlayer. This gives effective surface-to-surface distances of $D_{\rho}\simeq 43$, 125, and 210~nm for $N_{\mathrm{surf}}=22$, 11.5, and $7~\mu\mathrm{m}^{-2}$, respectively.

For a given probe distance, the temperature maps in Fig.~\ref{Janus_map_planar} are constructed from two steady-state temperature-increment fields extracted on two parallel planes (see upper insets in Fig.~\ref{Janus_map_planar}). The field $\Delta T_1(x,y)$ is evaluated on the plane passing through the upper probe points of all NPs, namely $A$, $B$, or $C$, whereas $\Delta T_2(x,y)$ is evaluated on the plane passing through the corresponding lower probe points, namely $A'$, $B'$, or $C'$. Thus, probe distances of 0, 85, and 170~nm correspond to the pairs $(A, A')$, $(B, B')$, and $(C, C')$ (as before), respectively. At each in-plane position $(x,y)$ within the circular domain, the local Janus factor is obtained by applying Eq.~\ref{def_facteur_Janus} to these two extracted fields.

For the probe planes at 0~nm, shown in the first column of Fig.~\ref{Janus_map_planar}, for the three concentrations (Fig.~\ref{Janus_map_planar}(a) for 22, Fig.~\ref{Janus_map_planar}(d) for 11.5, and Fig.~\ref{Janus_map_planar}(g) for 7), the maps exhibit pronounced and highly localized Janus hot spots centered on individual NPs. This highlights the near-field nature of the thermal asymmetry at the particle scale.

As the probe planes become farther away (to 85 and 170~nm, second and third columns of Fig.~\ref{Janus_map_planar}, respectively) the spatial distributions progressively homogenize: individual hot spots broaden and overlap, leading to smoother maps with reduced spatial contrast. This evolution reflects the progressive shift, discussed in Sec.~\ref{sec:nearest_neighbor}, from a response dominated by the screened self term $F_{J0}/(1+\alpha)$ (see e.g. blue solid line in Fig.~\ref{fig:weighted_terms_main}(b)), toward a regime where the collective contribution $F_{J}^{\mathrm{col}}/(1 + 1/\alpha)$ becomes increasingly important (see e.g. orange and green curves in Fig.~\ref{fig:weighted_terms_main}(b)). It also illustrates the main limitation of using Janus nanoheaters in dense assemblies under continuous-wave illumination: The collective background needed to increase the absolute temperature rise progressively homogenizes the local Janus contrast.

The maximum values in these maps follow the qualitative interparticle-distance dependence established in the dimer analysis (see e.g.\ Fig.~\ref{fig:weighted_terms_main}(a)). For the 0~nm probe distance, corresponding to $(A, A')$, the maps display larger maxima at lower densities (Fig.~\ref{Janus_map_planar}(a) vs Fig.~\ref{Janus_map_planar}(g)). For the 85~nm probe distance, corresponding to $(B, B')$, the response is non-monotonic with density. For the 170~nm probe distance, corresponding to $(C, C')$, larger maxima are obtained at higher densities. This confirms that the local Janus response in random arrays remains strongly governed by the observation distance and by the characteristic interparticle spacing.

From the maps of Fig.~\ref{Janus_map_planar}, it can be observed that the Janus factor of each NP depends on its precise location in the array. Indeed, the effective absorption cross section varies from one NP to another because of effects such as optical screening, leading to locally varying thermal influences on neighboring NPs. These maps therefore demonstrate that the Janus factor becomes an environment-dependent quantity once collective heating is present: identical NPs may exhibit substantially different thermal asymmetries depending on their position within the assembly.

More precisely, this spatial variation can be described by Eq.~\ref{eq:FJ_array_screening}: The Janus factor is reduced when the collective thermal background ($\Delta T_{\mathrm{col}}$) becomes large compared with the self-heating contribution of the probed particle ($\Delta T_{1}^{\mathrm{self}}+\Delta T_{2}^{\mathrm{self}}$). Since the collective background mainly scales with the array-averaged absorbed power, $I\left\langle \sigma_{\mathrm{abs}}\right\rangle_{\mathrm{NPs}}$, whereas the self contribution of particle $i$ scales with its own absorbed power, $I\sigma_{\mathrm{abs}}^{(i)}$, the screening strength is controlled, to first order, by the ratio $\left\langle \sigma_{\mathrm{abs}}\right\rangle_{\mathrm{NPs}}/\sigma_{\mathrm{abs}}^{(i)}$. Particles with $\sigma_{\mathrm{abs}}^{(i)}<\left\langle \sigma_{\mathrm{abs}}\right\rangle_{\mathrm{NPs}}$ are therefore more strongly screened, while particles with larger absorption cross sections ($\sigma_{\mathrm{abs}}^{(i)}>\left\langle \sigma_{\mathrm{abs}}\right\rangle_{\mathrm{NPs}}$) preserve a stronger local Janus contrast.

The remaining density dependence is governed by the characteristic interparticle distance. At low surface number density (about $N_{\mathrm{surf}}=7$ and below), neighboring NPs mainly generate a nearly symmetric collective background, leading to a screening-dominated regime (like in Eq.~\ref{eq:FJ_array_screening}). At higher $N_{\mathrm{surf}}$, interparticle distances become comparable to the characteristic thermal-Janus decay length (approximately 215~nm for the geometry considered here; see Fig.~\ref{Monomer_temp}), so short-range neighbor-induced asymmetries can partially compensate collective screening. As discussed in Sec.~\ref{sec:nearest_neighbor}, this near-field contribution is weak for $(A, A')$ but becomes more influential at larger observation distances, namely for $(B, B')$ and $(C, C')$ (see Fig.~\ref{fig:weighted_terms_main}(b), orange and green lines).

\subsection{Study of the collective Janus factor as a function of NP number and concentration}
\label{sec:array_averaged}

In this section, we investigate the evolution of the array Janus factor, $\left\langle F_J^{\mathrm{array}}\right\rangle_{\mathrm{NPs}}$ (only values over NP positions averaged), as a function of the surface number density ($N_{\mathrm{surf}}$) and of the total number of NPs in the array, denoted $N$. For each NP, the Janus factor (labeled $F_{J,i}^{\mathrm{array}}$) is evaluated from the temperature rises measured at the reference probe points $A$ and $A'$, located on opposite sides of the particle. Specifically, for an array containing $N$ NPs, we define
\begin{equation}
\left\langle F_J^{\mathrm{array}}\right\rangle_{\mathrm{NPs}}
=
\frac{1}{N}\sum_{i=1}^{N} F_{J,i}^{\mathrm{array}} .
\end{equation}

\begin{figure}[!ht]
  \centering
  \includegraphics[width=0.8\columnwidth]{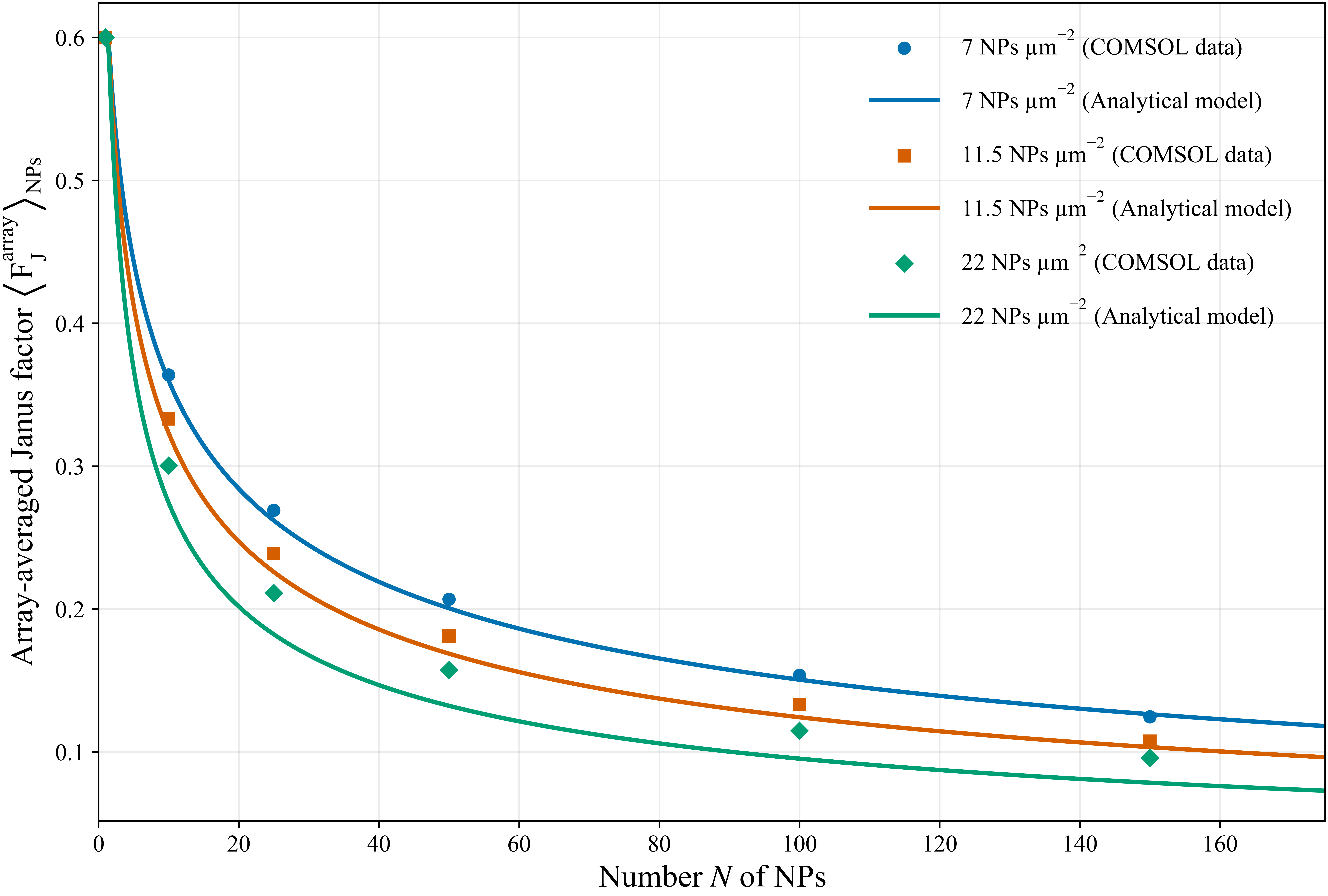}
  \caption{\label{Janus_vs_NPs}
  Array-averaged Janus factor $\left\langle F_J^{\mathrm{array}}\right\rangle_{\mathrm{NPs}}$ as a function of the number of NPs in the array, $N$, for $N_{\mathrm{surf}}=7$, 11.5, and $22~\mu\mathrm{m}^{-2}$. Discrete markers correspond to COMSOL simulations, while solid lines show the prediction of the analytical model given by Eq.~\ref{eq:model_final_FJ}.}
\end{figure}

The COMSOL simulation results are presented in Fig.~\ref{Janus_vs_NPs} (blue dots) for the three surface number densities considered above, $N_{\mathrm{surf}}=7$, 11.5, and $22~\mu\mathrm{m}^{-2}$, and for arrays containing between $N=10$ and 150 NPs. The arrays were generated following the procedure detailed in Sec.~\ref{sec:generation}.

The same overall trend is observed in all cases: $\left\langle F_J^{\mathrm{array}}\right\rangle_{\mathrm{NPs}}$ undergoes a rapid initial decay upon the addition of the first few neighbors, followed by a slower decrease as additional NPs are included (so as the array becomes larger). This behavior reflects the progressive build-up of a collective thermal background, which screens the intrinsic Janus response of each individual nanoheater. This trend is a quantitative measure of the homogenization of the Janus effect under continuous-wave illumination due to collective heating.

To describe this trend analytically, we use the screening-dominated approximation from Eq.~\ref{eq:FJ_array_screening}. In this limit, the collective contribution is treated as a nearly symmetric thermal background, and the array is modeled as a uniformly heated disk within the continuum approximation~\cite{baffou_thermoplasmonics_2017}. Since $F_{J,i}^{\mathrm{array}}$ is evaluated at the $N$ particle locations, and these positions are randomly generated within the disk, increasing $N$ makes the discrete average converge toward the spatial average of the temperature rise produced by a uniformly heated disk. The complete derivation, including the self-exclusion correction, is provided in Sec.~S2 of the Supporting Information. Under these assumptions, the closed-form model reads
\begin{align}
\label{eq:model_final_FJ}
\left\langle F_J^{\mathrm{array}}\right\rangle_{\mathrm{NPs}}
&\simeq
\frac{F_{J0}}{1+\dfrac{2\,\left\langle \Delta T_{\mathrm{col}}\right\rangle_{\mathrm{disk}}}{\left\langle S_0\right\rangle_{\mathrm{NPs}}}},
\\[6pt]
\label{eq:model_final_collective}
\left\langle \Delta T_{\mathrm{col}}\right\rangle_{\mathrm{disk}}
&\simeq
\frac{4\,I\,\left\langle \sigma_{\mathrm{abs}}\right\rangle_{\mathrm{NPs}}\,\sqrt{N_{\mathrm{surf}}}\,\sqrt{N}}
{3\,\pi^{3/2}\,\kappa_{\mathrm{water}}}
-
\frac{\sqrt{N_{\mathrm{surf}}}\,I\,\left\langle \sigma_{\mathrm{abs}}\right\rangle_{\mathrm{NPs}}}
{2\sqrt{\pi}\,\kappa_{\mathrm{water}}},
\\[6pt]
\label{eq:model_final_S0}
\left\langle S_0\right\rangle_{\mathrm{NPs}}
&=
\left\langle \Delta T_{1}^{\mathrm{self}}+\Delta T_{2}^{\mathrm{self}}\right\rangle_{\mathrm{NPs}}
=
\gamma\,I\,\left\langle \sigma_{\mathrm{abs}}\right\rangle_{\mathrm{NPs}},
\end{align}
where $F_{J0}$ is the isolated-NP Janus factor (as defined before) and $\gamma=S_0/(I\,\sigma_{\mathrm{abs}})$ is a proportionality constant extracted from a single-NP simulation, with $S_0=\Delta T_{1}^{\mathrm{self}}+\Delta T_{2}^{\mathrm{self}}$. In this framework, the collective contribution is evaluated using the continuum disk approximation (first term of $\left\langle \Delta T_{\mathrm{col}}\right\rangle_{\mathrm{disk}}$ in Eq.~\ref{eq:model_final_collective}), where the expression for the disk-averaged temperature rise is taken from Ref.~\onlinecite{pons_average_2023}. The second term (of Eq.~\ref{eq:model_final_collective}) corresponds to the self-contribution correction, following the standard self-exclusion procedure used for uniformly heated particle assemblies~\cite{Baffou2013}. 

A key advantage of Eq.~\ref{eq:model_final_FJ} is its simplicity. Since both $\left\langle S_0\right\rangle_{\mathrm{NPs}}$ and $\left\langle \Delta T_{\mathrm{col}}\right\rangle_{\mathrm{disk}}$ scale with $I\,\left\langle \sigma_{\mathrm{abs}}\right\rangle_{\mathrm{NPs}}$, the explicit evaluation of \(\left\langle \sigma_{\mathrm{abs}}\right\rangle_{\mathrm{NPs}}\) is not required to predict the array-averaged Janus factor. More explicitly, to determine the array Janus factor, one needs to simulate the Janus factor of a single particle, and simply put the values of $N$, $N_{\mathrm{surf}}$, and $\kappa_{\mathrm{water}}$ in Eqs.~\eqref{eq:model_final_FJ} to~\eqref{eq:model_final_S0}.

The solid lines in Fig.~\ref{Janus_vs_NPs} correspond to the prediction of Eq.~\ref{eq:model_final_FJ}. For the lowest surface number density, $N_{\mathrm{surf}}=7~\mu\mathrm{m}^{-2}$, the model provides a satisfactory description of the numerical data. This agreement is expected because the corresponding density-based surface-to-surface distance, $D_{\rho}=210$~nm, remains close to the characteristic thermal-Janus decay length of approximately $215$~nm, beyond which the single-particle Janus asymmetry becomes weak and the NPs effectively behave as point-like heat sources.

As the surface number density increases, the model systematically underestimates $\left\langle F_J^{\mathrm{array}}\right\rangle_{\mathrm{NPs}}$. This deviation is consistent with the emergence of short-range positive contributions that are not included in the pure collective-screening approximation. At higher densities, neighboring NPs are close enough to probe the finite spatial extent of the single-particle thermal asymmetry and to induce local asymmetric temperature contributions at the measurement points. These contributions partially compensate collective screening, resulting in larger simulated Janus factors than predicted by the pure collective-screening approximation. The discrepancy between the model and COMSOL data therefore provides a direct signature of the transition from a dilute, screening-dominated regime toward a denser regime where near-field neighbor-induced asymmetries become relevant. In addition, Sec.~S3 of the Supporting Information shows that the PDMS-mediated heat-flow redistribution becomes more pronounced as the surface number density increases. This behavior provides an additional signature of positive near-field effects in dense arrays, where short-range material-driven thermal asymmetries contribute to the deviations from the pure collective-screening approximation.

Furthermore, using Eq.~\ref{eq:model_final_FJ}, one can estimate the array size required to reach a target mean Janus factor. For example, solving Eq.~\ref{eq:model_final_FJ} for a target value $\left\langle F_J^{\mathrm{array}}\right\rangle_{\mathrm{NPs}}=0.20$ at a surface number density of $7~\mu\mathrm{m}^{-2}$ gives an array size of approximately $N\simeq 50$ NPs. To then determine specific temperature values, an optical simulation of the array is needed to evaluate $\left\langle \sigma_{\mathrm{abs}}\right\rangle_{\mathrm{NPs}}$, which is then used in Eqs.~\ref{eq:model_final_collective} and~\ref{eq:model_final_S0}. For example, considering an irradiance $I=1.155\times 10^{3}\ \mathrm{W\,cm^{-2}}$ and a mean absorption cross section $\left\langle \sigma_{\mathrm{abs}}\right\rangle_{\mathrm{NPs}} = 4.7\times 10^{-14}\ \mathrm{m^{2}}$, obtained for a planar array with 50 NPs at $\lambda=875$~nm, Eq.~\ref{eq:model_final_collective} gives $\left\langle \Delta T_{\mathrm{col}}\right\rangle_{\mathrm{disk}}\simeq 3.34$~K. The temperature rises at the two measurement points, namely $A$ and $A'$, are then reconstructed by rescaling the isolated-NP self-contributions consistently with Eq.~\ref{eq:model_final_S0}, using the same factor $I\left\langle \sigma_{\mathrm{abs}}\right\rangle_{\mathrm{NPs}}$. Adding the collective background obtained above gives reconstructed mean temperature rises of $\Delta T_1 \simeq 6~\mathrm{K}$ and $\Delta T_2 \simeq 4~\mathrm{K}$.

This example illustrates that, in principle, on average one side of the nanoheaters can exceed a phototherapy-relevant temperature increase of about $5~\mathrm{K}$ while the opposite side remains below this value, with the mean collective heating level staying below $5~\mathrm{K}$. However, this operating point would require an irradiance approximately $3.5\times 10^{3}$ times higher than the Maximum Permissible Exposure (MPE) for human skin. It is therefore unrealistic for \emph{in vivo} applications under standard safety constraints, but may remain relevant for controlled \emph{ex vivo} settings where larger irradiances can be considered. More broadly, directional heat release from Janus nanoheaters could also be useful for light-driven Janus micro/nanomotors based on self-thermophoretic propulsion~\cite{jiang_active_2010,liu_multiphysics_2023}, photocatalytic Janus heterostructures~\cite{pradhan_janus_2009}, solar steam generation and desalination~\cite{wang_facile_2023}, and solar-assisted extraction from seawater~\cite{fu_janus_2023}. This calculation underlines the central limitation identified throughout this work: In continuous-wave operation, increasing the number of particles helps to build up a therapeutically relevant temperature rise, but the associated collective background simultaneously homogenizes the temperature field and reduces the directional Janus contrast.

From a biomedical perspective, the present results reveal a fundamental design trade-off. Efficient photothermal therapy requires collective heating to reach biologically relevant temperature rises, whereas directional heat delivery requires the preservation of thermal asymmetry. Increasing the particle concentration favors the former objective but progressively suppresses the latter through thermal homogenization.

\section{Conclusion}

In this work, we quantified the thermo-Janus response of a core-shell Janus nanoheater consisting of a silica core, a gold shell, and an open PDMS overlayer, and investigated how this response evolves when nanoheaters are assembled into planar random arrays. At the single-particle level, we showed that the Janus effect is strongest in the near field and progressively vanishes at long distances, where the nanoheater behaves as an effective point-like heat source. By comparison with a reference case in which the PDMS thermal contrast is suppressed, $\kappa_{\mathrm{PDMS}}=\kappa_{\mathrm{water}}$, we further separated a geometrical contribution from a material-driven contribution associated with PDMS-mediated heat-flow redistribution.

Building on this characterization, we examined the role of close neighbors and identified two competing mechanisms: a nearly symmetric collective background temperature rise that screens the Janus factor, and short-range positive neighbor-induced contributions that emerge when interparticle distances become comparable to the spatial extent of the single-particle anisotropy. These trends were then extended to random arrays, where Janus maps revealed localized hot spots at short probe distances and a progressive homogenization at larger probe-plane distances.

Finally, we derived a closed-form analytical model for the array-averaged Janus factor $\left\langle F_J^{\mathrm{array}}\right\rangle_{\mathrm{NPs}}$ as a function of NP number and surface density. The model combines a continuum description of collective heating with a self-contribution correction and captures the evolution of $\left\langle F_J^{\mathrm{array}}\right\rangle_{\mathrm{NPs}}$ in the screening-dominated regime. At higher densities, systematic deviations from the model highlight the growing influence of short-range neighbor-induced asymmetries, which are not included in the pure collective-screening approximation.

Taken together, these results demonstrate that collective effects create a fundamental trade-off for Janus nanoheaters under continuous-wave illumination. On the one hand, increasing the number of NPs enhances the global temperature rise, which is necessary for biomedical photothermal applications. On the other hand, the same collective heating produces a nearly symmetric thermal background that progressively homogenizes the temperature field and screens the directional Janus response. For example, at a surface density of $N_{\mathrm{surf}}=7~\mu\mathrm{m}^{-2}$, increasing the array size to $N=50$ NPs reduces the array-averaged Janus factor from the isolated-particle value of about $0.60$ to approximately $0.21$, corresponding to a decrease of nearly $65\%$. The observable Janus behavior therefore reflects a balance between single-particle anisotropy, collective thermal screening, and short-range neighbor-induced asymmetries.

Direct biomedical implementation remains challenging under continuous-wave illumination conditions. The thermal Janus effect is strongest in the near field and under dilute conditions, but collective heating must be sufficiently strong to reach relevant temperature increments. As a result, the most favorable operating regime lies between two limits: very dilute systems, where directionality is preserved but heating is weak, and dense systems, where heating is enhanced but the Janus contrast is progressively erased by thermal homogenization. In the present simulations, taking $\left\langle F_J^{\mathrm{array}}\right\rangle_{\mathrm{NPs}}\gtrsim 0.2$ as a practical criterion for preserving a significant Janus response, this regime is mainly limited to arrays containing at most a few tens of NPs, typically $N\lesssim 25$ over the investigated density range, and up to about $N\simeq 50$ only for the most dilute case, $N_{\mathrm{surf}}=7~\mu\mathrm{m}^{-2}$.

Overall, our framework provides a practical route to predict the collective thermo-Janus behavior of random arrays from a single-particle thermoplasmonic simulation, enabling rapid exploration of density, array size, and nanoheater design. More generally, our results identify collective thermal screening as a central physical constraint on directional thermoplasmonic heating and establish quantitative scaling trends that can guide the future design of Janus nanoheater assemblies.

\section*{S1. Planar point-source model}

\begin{figure}[!ht]
    \centering
    \includegraphics[width=0.7\columnwidth]{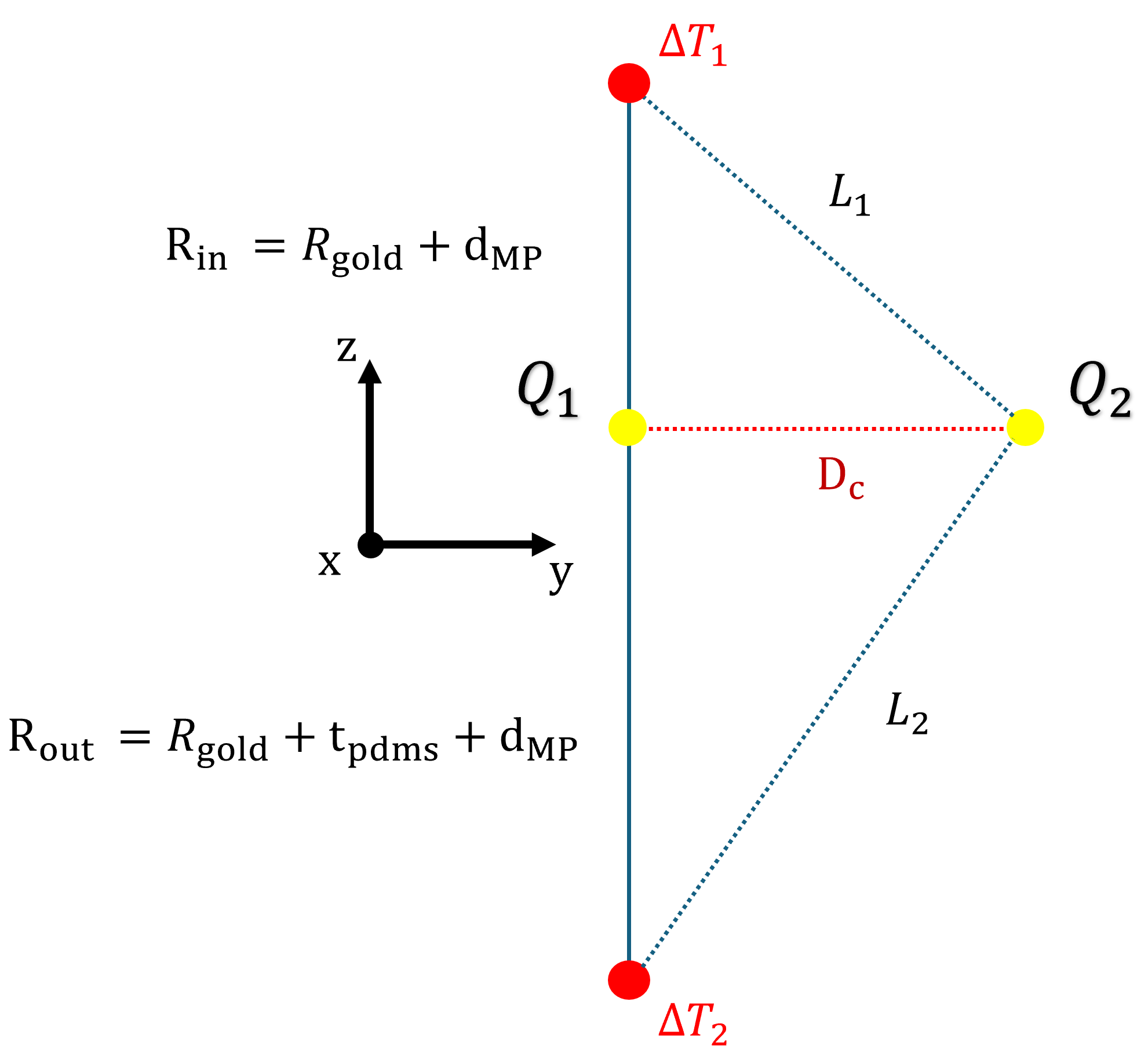}
    \caption{\label{fig:SI_model_planar}
    Planar point-source model used to rationalize the distance dependence of the apparent Janus response. The two probe points \(\Delta T_1\) and \(\Delta T_2\) are positioned at distances \(R_{\mathrm{in}}=R_{\mathrm{gold}}+d_{\mathrm{MP}}\) and \(R_{\mathrm{out}}=R_{\mathrm{gold}}+t_{\mathrm{PDMS}}+d_{\mathrm{MP}}\) from the reference source \(Q_1\). The surface-to-surface separation is denoted by \(D\), while the corresponding center-to-center distance between the point sources is \(D_{\mathrm{c}}=D+2R_{\mathrm{NP}}\). The distances from \(Q_2\) to the two probes are \(L_1\) and \(L_2\).}
\end{figure}

In this section, we apply the general decomposition derived in Eq.~(7) of the main text to a simplified planar two-source geometry, in order to separate the screened self contribution from the collective-asymmetry contribution to the Janus factor. The system is sketched in Fig.~\ref{fig:SI_model_planar}. Two steady point heat sources, of strengths \(Q_1\) and \(Q_2\), represent the probed NP and its nearest neighbor, respectively. The surface-to-surface separation between the two NPs is denoted by \(D\), consistently with the notation used in the main text. The corresponding center-to-center distance between the two point sources is therefore
\begin{equation}
D_{\mathrm{c}} = D + 2R_{\mathrm{NP}},
\end{equation}
where \(R_{\mathrm{NP}}=R_{\mathrm{gold}}+t_{\mathrm{PDMS}}\) is the effective outer radius of the Janus NP.

The Janus response is evaluated from two probe points, associated with the temperature rises \(\Delta T_1\) and \(\Delta T_2\), located along a line perpendicular to the \(Q_1\) to \(Q_2\) axis and passing through \(Q_1\). Their positions are determined by the gold radius \(R_{\mathrm{gold}}\), the PDMS thickness \(t_{\mathrm{PDMS}}\), and an additional measuring-point distance \(d_{\mathrm{MP}}\), which controls the probe-to-source standoff.

For convenience, we first recall the general expression established in the main text:
\begin{equation}
\label{eq:SI_general_decomposition}
F_J^{\mathrm{array}}
=
\frac{
\left(\Delta T_1^{\mathrm{self}}+\Delta T_1^{\mathrm{col}}\right)
-
\left(\Delta T_2^{\mathrm{self}}+\Delta T_2^{\mathrm{col}}\right)
}{
\left(\Delta T_1^{\mathrm{self}}+\Delta T_1^{\mathrm{col}}\right)
+
\left(\Delta T_2^{\mathrm{self}}+\Delta T_2^{\mathrm{col}}\right)
}
=
\frac{F_{J0}}{1+\alpha}
+
\frac{F_J^{\mathrm{coll}}}{1+1/\alpha},
\end{equation}
where
\begin{equation}
F_{J0}
=
\frac{\Delta T_1^{\mathrm{self}}-\Delta T_2^{\mathrm{self}}}
{\Delta T_1^{\mathrm{self}}+\Delta T_2^{\mathrm{self}}},
\qquad
F_J^{\mathrm{coll}}
=
\frac{\Delta T_1^{\mathrm{col}}-\Delta T_2^{\mathrm{col}}}
{\Delta T_1^{\mathrm{col}}+\Delta T_2^{\mathrm{col}}},
\qquad
\alpha
=
\frac{\Delta T_1^{\mathrm{col}}+\Delta T_2^{\mathrm{col}}}
{\Delta T_1^{\mathrm{self}}+\Delta T_2^{\mathrm{self}}}.
\end{equation}

We now specify these quantities for the planar point-source geometry of Fig.~\ref{fig:SI_model_planar}. The probe positions relative to \(Q_1\) are
\[
R_{\mathrm{in}}=R_{\mathrm{gold}}+d_{\mathrm{MP}},
\qquad
R_{\mathrm{out}}=R_{\mathrm{gold}}+t_{\mathrm{PDMS}}+d_{\mathrm{MP}},
\]
while the distances from \(Q_2\) to the two probes are
\begin{equation}
\label{eq:SI_L1L2_planar}
L_1=\sqrt{D_{\mathrm{c}}^2+R_{\mathrm{in}}^2},
\qquad
L_2=\sqrt{D_{\mathrm{c}}^2+R_{\mathrm{out}}^2}.
\end{equation}

In a homogeneous medium of thermal conductivity \(\kappa\), the steady Green's function is
\begin{equation}
G(r)=\frac{1}{4\pi\kappa r}.
\end{equation}
The four temperature contributions entering Eq.~\eqref{eq:SI_general_decomposition} are therefore
\begin{equation}
\Delta T_1^{\mathrm{self}} = Q_1\,G(R_{\mathrm{in}}),
\qquad
\Delta T_2^{\mathrm{self}} = Q_1\,G(R_{\mathrm{out}}),
\end{equation}
and
\begin{equation}
\Delta T_1^{\mathrm{col}} = Q_2\,G(L_1),
\qquad
\Delta T_2^{\mathrm{col}} = Q_2\,G(L_2).
\end{equation}

Using Eq.~\eqref{eq:SI_L1L2_planar}, the collective Janus contribution and the weighting factor read
\begin{equation}
\label{eq:SI_FJcoll_alpha_explicit}
F_J^{\mathrm{coll}}(D,d_{\mathrm{MP}})
=
\frac{
\dfrac{1}{L_1}-\dfrac{1}{L_2}
}{
\dfrac{1}{L_1}+\dfrac{1}{L_2}
},
\qquad
\alpha(D,d_{\mathrm{MP}})
=
a\,
\frac{
\dfrac{1}{L_1}+\dfrac{1}{L_2}
}{
\dfrac{1}{R_{\mathrm{in}}}+\dfrac{1}{R_{\mathrm{out}}}
},
\end{equation}
with \(a\equiv Q_2/Q_1\).

Combining the self and collective terms, the apparent Janus factor predicted by the planar point-source model becomes
\begin{equation}
\label{eq:SI_planar_explicit}
F_J^{\mathrm{array}}
=
\frac{
\left(\dfrac{1}{R_{\mathrm{in}}}-\dfrac{1}{R_{\mathrm{out}}}\right)
+
a\left(\dfrac{1}{L_1}-\dfrac{1}{L_2}\right)
}{
\left(\dfrac{1}{R_{\mathrm{in}}}+\dfrac{1}{R_{\mathrm{out}}}\right)
+
a\left(\dfrac{1}{L_1}+\dfrac{1}{L_2}\right)
}.
\end{equation}

\section*{S2. Complete derivation of the closed-form model for the array-averaged Janus factor}

We now derive a closed-form expression to predict \(\left\langle F_J^{\mathrm{array}}\right\rangle_{\mathrm{NPs}}\) using only single-particle Janus simulations. We assume that the system operates in the low-density regime, where the Janus response is primarily governed by collective screening. Starting from the screening form of Eq.~(8) of the main text and averaging over NPs, we obtain
\begin{equation}
\label{eq:Jarray_mean_0th}
\left\langle F_J^{\mathrm{array}}\right\rangle_{\mathrm{NPs}}
\simeq
\frac{F_{J0}}
{1+2\,\left\langle \Delta T_{\mathrm{collective}}\right\rangle_{\mathrm{NPs}}/\left\langle S_0\right\rangle_{\mathrm{NPs}}},
\end{equation}
where \(S_0=\Delta T_1^{\mathrm{self}}+\Delta T_2^{\mathrm{self}}\). This expression corresponds to retaining only the zeroth-order term of the averaging procedure. To assess this approximation, we evaluated the zeroth-order prediction against the full point-source average over the same random particle configurations used in the simulations. For the array sizes and densities considered here, the relative deviation remained small ($<1\%$) compared with the deviations associated with near-field contributions, justifying the use of the zeroth-order term in the closed-form model.

We now employ the continuum approximation~\cite{baffou_thermoplasmonics_2017}, modeling the array as a uniformly heated disk. Since \(F_J^{\mathrm{array}}\) is evaluated at the \(N\) particle locations, and these positions are randomly generated within the disk, increasing \(N\) makes the discrete average converge toward the spatial average of the temperature rise produced by a uniformly heated disk. We therefore replace \(\left\langle \Delta T_{\mathrm{collective}}\right\rangle_{\mathrm{NPs}}\) by the disk-averaged value \(\left\langle \Delta T_{\mathrm{collective}}\right\rangle_{\mathrm{disk}}\) in Eq.~\eqref{eq:Jarray_mean_0th}.

A key difficulty is that we need the collective contribution, namely the temperature rise generated by all NPs of the array except the one located at the observation point. We write
\begin{equation}
\label{eq:Tcollective_disk_def}
\left\langle \Delta T_{\mathrm{collective}}\right\rangle_{\mathrm{disk}}
=
\left\langle \Delta T_{\mathrm{all}}\right\rangle_{\mathrm{disk}}
-
\left\langle \Delta T_{\mathrm{self}}\right\rangle_{\mathrm{disk}},
\end{equation}
where \(\left\langle \Delta T_{\mathrm{all}}\right\rangle_{\mathrm{disk}}\) denotes the disk-averaged temperature rise produced by the full uniformly heated disk, and \(\left\langle \Delta T_{\mathrm{self}}\right\rangle_{\mathrm{disk}}\) corresponds to the averaged self contribution that must be removed.

The disk-averaged temperature rise produced by the uniformly heated disk admits the closed-form expression~\cite{pons_average_2023}
\begin{equation}
\label{eq:Tall_disk_closedform}
\left\langle \Delta T_{\mathrm{all}}\right\rangle_{\mathrm{disk}}
=
\frac{4\,I\,\left\langle\sigma_{\mathrm{abs}}\right\rangle_{\mathrm{NPs}}\,\sqrt{N_{\mathrm{surf}}}\,\sqrt{N}}
{3\,\pi^{3/2}\,\kappa_{\mathrm{water}}},
\end{equation}
where \(I\) is the incident irradiance, \(\left\langle\sigma_{\mathrm{abs}}\right\rangle_{\mathrm{NPs}}\) is the array-averaged absorption cross section, \(N_{\mathrm{surf}}\) is the surface number density, \(N\) is the total number of NPs, and \(\kappa_{\mathrm{water}}\) is the thermal conductivity of the embedding medium.

To compute the collective contribution, we subtract an averaged self term. In the continuum disk picture, the standard self-exclusion correction consists in removing, around each observation point, a circular region whose area equals the average area associated with one NP in the array~\cite{Baffou2013}. A closed-form self-exclusion correction can be obtained when evaluating the temperature rise at a fixed observation point such as the disk center, since the exclusion region is fully contained within the heated domain and the geometry remains radially symmetric. In our case, however, \(\left\langle \Delta T_{\mathrm{self}}\right\rangle_{\mathrm{disk}}\) is defined as a spatial average over observation points distributed across the finite disk. This average includes points located near the disk boundary, for which the exclusion region is truncated by the disk edge. In principle, the self contribution should therefore be computed with a position-dependent truncated exclusion area and then averaged over the disk. Numerical checks indicate that the corresponding correction is small for the array sizes and densities considered here. We therefore neglect boundary truncation and use the closed-form approximation
\begin{equation}
\label{eq:Tself_disk_closedform}
\left\langle \Delta T_{\mathrm{self}}\right\rangle_{\mathrm{disk}}
\simeq
\frac{\sqrt{N_{\mathrm{surf}}}\, I\, \left\langle \sigma_{\mathrm{abs}}\right\rangle_{\mathrm{NPs}}}
{2\sqrt{\pi}\,\kappa_{\mathrm{water}}}.
\end{equation}

This yields the following model:
\begin{equation}
\label{eq:SI_model_final}
\begin{aligned}
\left\langle F_J^{\mathrm{array}}\right\rangle_{\mathrm{NPs}}
&\simeq
\frac{F_{J0}}
{1+\dfrac{2\,\left\langle \Delta T_{\mathrm{collective}}\right\rangle_{\mathrm{disk}}}{\left\langle S_0\right\rangle_{\mathrm{NPs}}}},
\\[6pt]
\left\langle \Delta T_{\mathrm{collective}}\right\rangle_{\mathrm{disk}}
&=
\frac{4\,I\,\left\langle\sigma_{\mathrm{abs}}\right\rangle_{\mathrm{NPs}}\,\sqrt{N_{\mathrm{surf}}}\,\sqrt{N}}
{3\,\pi^{3/2}\,\kappa_{\mathrm{water}}}
-
\frac{\sqrt{N_{\mathrm{surf}}}\,I\,\left\langle \sigma_{\mathrm{abs}}\right\rangle_{\mathrm{NPs}}}
{2\sqrt{\pi}\,\kappa_{\mathrm{water}}},
\\[6pt]
\left\langle S_0\right\rangle_{\mathrm{NPs}}
&=
\left\langle \Delta T_1^{\mathrm{self}}+\Delta T_2^{\mathrm{self}}\right\rangle_{\mathrm{NPs}}
=
\gamma\, I\,\left\langle \sigma_{\mathrm{abs}}\right\rangle_{\mathrm{NPs}}.
\end{aligned}
\end{equation}
Here, \(F_{J0}\) is the isolated-particle Janus factor and \(\gamma=S_0/(I\sigma_{\mathrm{abs}})\) is a proportionality constant obtained from a single-NP simulation. Since both \(\left\langle S_0\right\rangle_{\mathrm{NPs}}\) and \(\left\langle \Delta T_{\mathrm{collective}}\right\rangle_{\mathrm{disk}}\) scale with \(I\left\langle\sigma_{\mathrm{abs}}\right\rangle_{\mathrm{NPs}}\), the explicit evaluation of \(\left\langle\sigma_{\mathrm{abs}}\right\rangle_{\mathrm{NPs}}\) is not required to operate the model. In practice, it is sufficient to run a single-NP COMSOL simulation using an arbitrary irradiance \(I\) and absorption cross section to extract \(F_{J0}\) and \(\gamma\), and to use these values consistently in Eq.~\eqref{eq:SI_model_final}.

\section*{S3. PDMS influence}

To further isolate the role of geometry from material-driven effects, we recast the analysis in terms of a geometrical Janus map, in direct analogy with the geometrical Janus factor introduced for the isolated NP. Specifically, we define the geometrical array response as the Janus map obtained when suppressing the PDMS thermal contrast by imposing \(\kappa_{\mathrm{PDMS}}=\kappa_{\mathrm{water}}\), namely
\begin{equation}
F_{J,\mathrm{geom}}^{\mathrm{array}}(x,y)
=
\left.
\frac{\Delta T_1(x,y)-\Delta T_2(x,y)}
{\Delta T_1(x,y)+\Delta T_2(x,y)}
\right|_{\kappa_{\mathrm{PDMS}}=\kappa_{\mathrm{water}}},
\end{equation}
which captures the asymmetry arising solely from the geometrical definition of the probe points and the spatial distribution of the temperature field, without PDMS-mediated heat-flux redistribution. The full material-inclusive Janus map is
\begin{equation}
F_J^{\mathrm{array}}(x,y)
=
\left.
\frac{\Delta T_1(x,y)-\Delta T_2(x,y)}
{\Delta T_1(x,y)+\Delta T_2(x,y)}
\right|_{\kappa_{\mathrm{PDMS}}}.
\end{equation}

We then quantify the PDMS-mediated contribution by comparing \(F_J^{\mathrm{array}}\) to its geometrical counterpart. A convenient pointwise metric is the relative difference map
\begin{equation}
\eta_{\kappa}(x,y)
=
\left(
\frac{
F_J^{\mathrm{array}}(x,y)
-
F_{J,\mathrm{geom}}^{\mathrm{array}}(x,y)
}
{F_J^{\mathrm{array}}(x,y)}
\right)\times 100,
\end{equation}
which encodes the fraction of the local Janus response related to PDMS-driven heat-flow redistribution.

\begin{figure}[!ht]
 \centering
 \includegraphics[width=0.7\columnwidth]{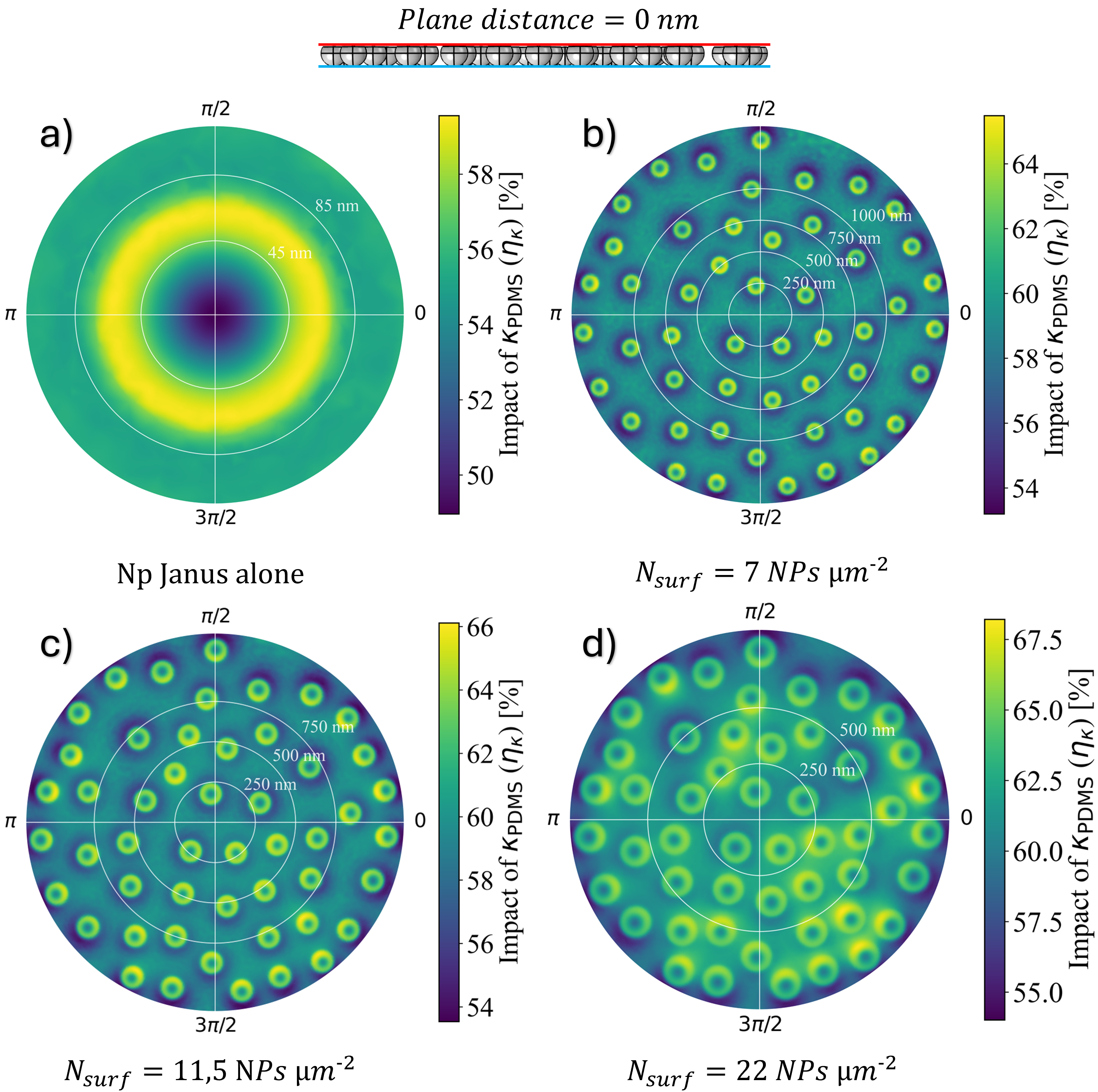}
 \caption{\label{PDMSimpact}
 Impact of the PDMS thermal conductivity on the local Janus response at a probe distance of 0~nm. Top inset: side view of the probe planes coincident with the NP layer; red and blue lines denote the two probe planes used to compute the Janus factor. Polar maps report the relative PDMS-mediated contribution defined as \(\eta_{\kappa}(x,y)=\left[F_J^{\mathrm{array}}(x,y)-F_{J,\mathrm{geom}}^{\mathrm{array}}(x,y)\right]/F_J^{\mathrm{array}}(x,y)\times 100\), where \(F_J^{\mathrm{array}}\) is the full Janus map and \(F_{J,\mathrm{geom}}^{\mathrm{array}}\) is the geometrical Janus map obtained by imposing \(\kappa_{\mathrm{PDMS}}=\kappa_{\mathrm{water}}\). (a) Single isolated Janus NP. (b) to (d) Janus NP arrays with \(N_{\mathrm{surf}}=7\), 11.5, and \(22~\mathrm{NPs}\,\mu\mathrm{m}^{-2}\), respectively.}
\end{figure}

Fig.~\ref{PDMSimpact} reports \(\eta_{\kappa}(x,y)\) for a probe distance of 0~nm, corresponding to the first column of Fig.~6 of the main text. For the isolated NP, the map yields at the particle center a contribution of about 50\%, consistent with the single-particle analysis of Fig.~3 of the main text. In the array case, the magnitude and spatial extent of \(\eta_{\kappa}\) increase with \(N_{\mathrm{surf}}\), indicating that the PDMS-mediated contribution becomes progressively more influential as interparticle distances decrease and thermal cross-talk strengthens. This trend supports the interpretation that, in dense arrays, PDMS-driven heat-flux redistribution enhances short-range neighbor-induced asymmetries and thereby promotes positive contributions to the measured Janus response.
 
\bibliography{aipsamp}

\subsection*{Funding}

This work was supported by the Action de Recherche Concertée (project ARC-23/27 UMONS3).

This work was supported by the Fonds de la Recherche Scientifique – FNRS under Grant No.40034930.

\end{document}